\documentclass[a4paper,11pt]{article}
\pdfoutput=1 

\usepackage{jcappub} 

\usepackage[T1]{fontenc} 
\usepackage{amsmath}
\usepackage{amsfonts}
\usepackage{amsthm}
\usepackage{amssymb}
\usepackage{latexsym, array,multirow,  verbatim, enumerate}
\usepackage{slashed}
\usepackage{booktabs}
\usepackage{tabularx}
\usepackage{cancel}
\usepackage{dcolumn}
\usepackage{bm}
\usepackage{graphicx, caption, subcaption} 
\usepackage{url}
\usepackage[compat=1.0.0]{tikz-feynman}
\usepackage{pifont}
\usepackage{empheq}
\usepackage[utf8]{inputenc}
\usepackage{bm}
\usepackage{blkarray}
\usepackage[font=footnotesize,labelfont=sf]{caption}
\usepackage{cleveref}
\usepackage{float}
\usepackage{multirow}
\usepackage{physics}
\graphicspath{{./Figures/}}
\usepackage{tabularx}
\usepackage{ulem}

\newcommand{\C}{\textcolor{cyan}}
\newcommand{\xvec}{\vec{x}}

\newcommand{\Cops}[2]{\mathbb{C}^{(#1)}_{#2}}
\newcommand{\boldk}{\boldsymbol{k}}
\newcommand{\boldx}{\boldsymbol{x}}

\title{ Time Non-locality in Dark Matter and LSS }

\author[a]{Arhum Ansari,}
\author[a]{Arka Banerjee,}
\author[a]{Sachin Jain,}
\author[b]{and Shaunak Padhyegurjar}

\affiliation[a]{ Department of Physics, Indian Institute of Science Education and Research Pune, Pune 411008, India}
\affiliation[b]{ Department of Physics, Indian Institute of Science Education and Research Bhopal, Bhopal 462066, India}
\emailAdd{ansari.arhum@students.iiserpune.ac.in}
\emailAdd{arka@iiserpune.ac.in}
\emailAdd{sachin.jain@iiserpune.ac.in}
\emailAdd{shaunak19@iiserb.ac.in}

\abstract{ We explore the intriguing phenomenon of time non-locality in the evolution of dark matter and Large Scale Structure (LSS). Recently in\,\cite{Donath:2023sav}, it was shown that time non-locality emerges in bias tracer fluctuations, which are $SO(3)$ scalars in real space, at fifth order in the perturbation expansion in dark matter overdensity. We demonstrate that by breaking the symmetry down to $SO(2)$, which is the case whenever line-of-sight effects become important, such as for flux fluctuations in the Lyman $\alpha$ forest, the temporal non-locality appears at the third order in expansion. Additionally, within the framework of EFTofLSS, we demonstrate that time non-locality manifests in the effective stress tensor of dark matter, which is a second rank tensor under $SO(3)$ transformations, again at the third order in dark matter overdensity. Furthermore, we highlight the effectiveness of the standard $\Pi$ basis\,\cite{Mirbabayi:2014zca} in handling time non-local operators.} 

\begin{document}
\maketitle
\flushbottom


\section{Introduction}\label{sec:intro}

The Standard Model of cosmology includes an inflationary phase of accelerated expansion, leading to homogeneity and isotropy on large scales, while at the same time giving rise to small inhomogeneities from quantum fluctuations. These inhomogeneities are reflected in the anisotropy of the cosmic microwave background (CMB)\,\cite{WMAP:2012nax,Planck:2018vyg,Planck:2018jri} , and also serve as initial density perturbations that grow during matter domination era and give rise to large scale structures (LSS). The evolution of LSS perturbations, therefore, contains information about the initial spectrum of fluctuations but is also sensitive to the relative ratios and properties of energy components that drive the subsequent expansion, including Dark Energy (DE) and Dark Matter (DM). Understanding the distribution of LSS is fundamental to our understanding of very early Universe physics, as well to discern the ultimate fate of the Universe.

Due to the three-dimensional nature, the number of observable modes in LSS is larger compared to the CMB and therefore is, in principle, statistically more constraining on the parameters of interest. However, since gravitational evolution drives the perturbations into the nonlinear regime on small scales, accurate theoretical modeling down to these scales remains a challenge. The equations governing the dynamics of dark matter are the Boltzmann equations in an expanding background. One approach has been to use computationally intensive $N$-body simulations\,\cite{Angulo:2021kes,hockney2021computer,Springel:2006vs,Vogelsberger:2019ynw,Chisari:2011iq,Adamek:2013wja,Adamek:2016zes,Adamek:2017grt,Ali-Haimoud:2012fzp,Angulo:2016hjd,Bayer:2020tko,Chudaykin:2020hbf,Foreman:2015lca} of the dark matter phase space, under the assumption that DM is the major clustering component, subsequently driving the evolution of the baryon clustering. These studies have been extended beyond $\Lambda$CDM, see for e.g.\,\cite{Banerjee:2016zaa,Banerjee:2018bxy,Banerjee:2019bjp}. In a (hypothetical) Universe with DM as the only clustering component, these simulations are accurate down to the scales much smaller than the highly nonlinear, virialized halo structures.

The second approach to treating gravitational clustering beyond linear order is to use higher-order perturbation theory. In the Standard Perturbation Theory\,\cite{Bouchet:1992xz,Catelan:1994kt,Matsubara:1995kq,Bouchet:1994xp,Scoccimarro:2000sn,Bernardeau:2001qr}, the relevant equations are solved order by order using a series ansatz. However, even for quasilinear wavenumbers where the perturbation expansion should be technically valid, this approach is unable to reproduce the matter power spectrum computed through simulations. Therefore, principles of Effective Field Theories (EFT) have been applied to the context of the LSS perturbations to model the quasi-linear scales\,\cite{Baumann:2010tm,Carrasco:2012cv,Porto:2013qua,Senatore:2014via,Carrasco:2013mua,Angulo:2014tfa,Baldauf:2014qfa,Konstandin:2019bay,Bertolini:2016bmt,Abolhasani:2015mra}. In this approach, termed the EFT of LSS (EFTofLSS), we integrate out the short scale modes in SPT. The process of integrating out short modes generates an effective stress tensor in the equations of motion. The stress tensor incorporates all the short-scale physics which is needed to describe the distribution of dark matter at quasi-linear scales. Various correlation functions of the density fluctuations can then be calculated within the framework. Predictions of EFTofLSS have been benchmarked against full simulations to a high degree of accuracy on quasilinear scales, and these techniques have now been applied to various datasets to extract state-of-the-art constraints on various cosmological parameters\,\cite{Cabass:2022wjy,Cabass:2022ymb,Philcox:2021kcw,Ivanov:2021kcd,DAmico:2019fhj,Braganca:2023pcp,DAmico:2022osl,DAmico:2022gki,Zhang:2021yna,DAmico:2020tty,DAmico:2020kxu,Nishimichi:2020tvu,Colas:2019ret,Ivanov:2023qzb,Schmidt:2020viy}.

In the perturbative approach, baryonic matter such as galaxies are assumed to be tracers of the underlying dark matter field. The overdensity for tracers is given as a ``bias expansion"\,\cite{Desjacques:2016bnm}, in terms of the dark matter fields. In most cases, the expansion is assumed to be local in time - i.e. the overdensity of tracers at some time $t$ is written in terms of dark matter operators at the same time. However, it has been argued that since galaxies are not formed instantaneously --- the formation time is of order $(1/10)^{\text{th}}$ of the evolution timescales of the dark matter field itself --- the bias expansion should be non-local in time \cite{Senatore:2014eva}. The "local in time" and "non-local in time" terminology used here has been used in various published works\,\cite{Donath:2023sav,Ivanov:2023yla,Carroll:2013oxa,Angulo:2015eqa,DAmico:2022ukl}. Hence, we use the same terminology to be consistent with the literature. A more formal definition of time local and time non- local operators will be given in later sections. Recent investigations\,\cite{Donath:2023sav} have shown that, indeed, time non-locality is an essential feature of the bias expansion at the fifth order. Technically, it was shown that when writing a bias expansion consisting of zero derivative $SO(3)$ invariant operators of dark matter fields in real space, the local and the non-local basis differ at the fifth order in the perturbative expansion. This indicates that the physics of biased tracers cannot be described accurately by local expansion beyond the fourth order in perturbation theory.

While the bias expansion for the clustering of tracers in real space is written in terms of $SO(3)$ invariant operators, there are observationally relevant cases in cosmology when additional $SO(2)$ invariant terms must also be included in the expansion. Such cases typically arise when the line-of-sight (LoS) direction becomes special, therefore breaking the $SO(3)$ symmetry. This is exactly what happens while considering the clustering of tracers in redshift space \cite{Desjacques:2018pfv}  --- the main observable of spectroscopic surveys such as DESI\,\cite{DESI:2024mwx}. Another case where such operators are essential is in the bias expansion of the flux fluctuations in the Lyman-$
\alpha$ forest\,\cite{Ivanov:2023yla}. It is, therefore, interesting to investigate the issue of time non-locality for such $SO(2)$ invariant operators.

Time non-locality in an inherent part of the EFT framework for the reasons discussed above. This has motivated a lot of studies in this direction, see for example\,\cite{Abolhasani:2015mra,Senatore:2014eva,Ivanov:2023yla,Carroll:2013oxa,Angulo:2015eqa,DAmico:2022ukl,Mirbabayi:2014zca,Desjacques:2016bnm} and references therein. As stated before, most recent study in this direction investigated $SO(3)$ invariant operators and found that time non-locality manifests itself at fifth order for such operators\,\cite{Donath:2023sav}. Ours is first study which investigates time non-locality in the context of $SO(2)$ invariant operators, and we find that time non-locality appears at the third order for these operators, as opposed to the fifth order for the $SO(3)$ invariant operators. This implies that time non-locality may be detected in the 1-loop power spectrum for observables for which selection operators need to be included. In contrast, for the $SO(3)$ bias expansion, time non-locality can only be detected at the two-loop power spectrum of the observables.

Another question that arises is whether the issue of time non-locality plays any role in the evolution of the dark matter field itself within the context of EFTofLSS. Indeed it is the case with EFTofLSS and has been discussed in the literature\,\cite{Abolhasani:2015mra,Mirbabayi:2014zca,Desjacques:2016bnm}. Let us briefly explain the reason behind time non-locality in EFTofLSS. For the EFT approach to work, a separation of scales is needed between the scales at which we want predictions (i.e. the quasilinear scales) and the scales at which new (nonlinear) physics kicks in. The wavenumber at which the density perturbations become of $\mathcal{O}$(1) is known as the nonlinear wavenumber  ($k_{\text{NL}}$). This, therefore, defines a natural scale in the space domain and makes it possible to define an effective field theory (EFTofLSS), by integrating out the modes above the non-linear scale. The presence of a scale separation in space ensures that integrating out the short modes gives a space local effective theory. The local EFT has a derivative expansion that is regulated by the non-linear scale ($k_{\text{NL}}$). However, the typical timescales associated with the evolution of both the large scales and the small scales, are similar $\mathcal O(H^{-1})$. Therefore, interestingly, there is no scale separation in the time domain. Therefore, integrating out the short modes generically yields a time non-local EFT. This time non-locality manifests itself in the governing equations of LSS through the effective stress tensor in the equations of motion for dark matter\,\cite{Carrasco:2013mua,Carroll:2013oxa,Angulo:2014tfa,Baldauf:2014qfa,Bertolini:2016bmt}. The non-locality in stress tensor is because there is no hierarchy in evolution timescales between the long and the short modes. We have investigated the structure of the time non-local stress tensor beyond linear order. The time local expansions offer a simpler, and physically intuitive framework. Given its simplicity, and the fact that the two expansions match at the leading orders, the time local expansion is a useful tool for practical calculations. Our objective in this work is to point out up to what order in perturbation theory one can write time local counterterms in the effective stress tensor and still be robust in the analysis. We have found that the time non-local stress tensor differs from the time local stress tensor at third order. This is due to the tensorial nature of the stress tensor $\tau_{ij}$ ---- for scalars such as $\partial_i\partial_j\tau_{ij}$ the difference between the local-in-time and non-local-in-time expressions is at the fourth order, see Appendix.\,\ref{ap:scal_ops}.

We have demonstrated that the relevance of time non-local effects at a given order depends on the specific observable under consideration. For instance, when examining observables that depend solely on $SO(3)$ invariant operators in the bias expansion, time non-local effects manifest at fifth order\,\cite{Donath:2023sav}. Conversely, for observables that also depend on projections of dark matter fields (i.e.
$SO(2)$invariant structures), time non-local effects become significant at third order. Clarifying this distinction is also an objective of our analysis.

The paper is organised as follows. In Sec.\,\ref{sec:bg}, we review the basic equations of SPT and biased tracers relevant to our discussion. We describe an alternate system ($\Pi$-system) of writing operators in the bias expansion and also discuss the structure of selection operators. We use this section to also set up the conventions and notations that will be used throughout the paper. In Sec.\,\ref{sec:nloc_dm}, we discuss the EFTofLSS and the effective stress tensor. Then, we discuss how time non-locality appears in dark matter and lay out the details of getting the non-local stress tensor up to third order. Then, in Sec.\,\ref{sec:nloc_bias_exp}, we discuss time non-locality in the bias expansion of $SO(3)$ invariant operators. We also discuss the invertible mapping between the non-local and $\Pi$-basis. In Sec.\,\ref{sec:nloc_sel_op}, we discuss bias expansion with the $SO(2)$ invariant (selection) operators and the time non-local basis upto fourth order. Finally, in Sec.\,\ref{sec:disc_conc}, we conclude by stating our results and giving some future directions. In  appendices\,\ref{ap:vec_op} and \,\ref{ap:renorm} we discuss  the vector operator structure for dark matter and renormalisation of the flux field in Lyman $\alpha$ respectively.
\section{Background details}\label{sec:bg}

In this section, we will discuss the fluid equations governing the dynamics of dark matter at large scales. We will also discuss the SPT approach to solve the fluid equations (Sec.\ref{ssec:SPT}). We will then discuss how tracers (galaxies) are modelled and will describe the bias expansion for $SO(3)$ (Sec.\ref{ssec:loc_bias_exp}) and $SO(2)$ (Sec.\ref{ssec:sel_op}) invariant operators. In the end of this section, we will describe an alternate system ($\Pi$ system) for writing operators in the bias expansion which will be relevant for later discussion on time non-locality.

\subsection{Dark-Matter: Standard Perturbation Theory (SPT)}\label{ssec:SPT}
Dark matter at large scales has been shown to behave as an effective fluid following the \textit{Vlasov-Poisson} equation which is the dynamical equation describing collisionless matter coupled with gravity\,\cite{Baumann:2010tm,Carrasco:2012cv}. In the single stream approximation i.e. for a vanishing velocity dispersion, the Boltzmann hierarchy can be truncated after the first moment and we end up with the continuity and Euler equation for the dark matter overdensity and velocity field. The continuity and Euler equation for collisionless fluid is given as
\begin{align}\label{eq:fluid_eq}
    \delta'(\xvec,\eta) &+ \nabla\cdot[\vec{v}(\xvec,\eta)(\delta(\xvec,\eta) + 1)] = 0\;,\nonumber\\
    v_i'(\xvec,\eta) + &\mathcal{H} v_i(\xvec,\eta) + (\vec{v}(\xvec,\eta).\nabla)v_i(\xvec,\eta) = -\nabla_i\Phi(\xvec,\eta)\;,
\end{align}
where $^\prime$ denotes derivative w.r.t the conformal time $\eta$ and $\delta\equiv (\rho-\Bar{\rho})/\Bar{\rho}$ is the dark matter overdensity and $\Vec{v}$ is the velocity field. Here $\rho$ and $\Bar{\rho}$ denote the matter density and it's average respectively. The system of equations given in Eq.\eqref{eq:fluid_eq} along with the Poisson equation for gravitational potential
\begin{align}\label{eq:poisson_eq}
    \nabla^2\Phi = \frac{3}{2}\Omega_m(\eta)\mathcal{H}^2(\eta)\delta(\vec{x}, \eta)\;,
\end{align}
closes the system. Here $\mathcal{H}$ is the Hubble parameter expressed in conformal time. In the standard perturbation theory (SPT), the system of equations in Eq.\eqref{eq:fluid_eq} is solved by a series ansatz in momentum space as,
\begin{align}\label{eq:ansatz_series}
    \delta(k,\eta)=\sum_{n=1}^\infty D^n(\eta)\delta^{(n)}(k)\;,\nonumber\\
    \theta(k,\eta)=-\mathcal{H}f\sum_{n=1}^\infty D^n(\eta)\theta^{(n)}(k)\;,
\end{align}
where $D(\eta)$ is the solution of the linear equation for $\delta$ given in Eq.\eqref{eq:fluid_eq} and $f\equiv (d\ln D/d\ln a)$ is the logarithmic growth factor. At each order $\delta^{(n)}(k)$ and $\theta^{(n)}(k)$ are defined as a convolution over linear overdensity in momentum space,
\begin{align}\label{eq:kernels}
    \delta^{(n)}(\vec{k})=\int_{\vec{q}_1...\vec{q}_n}(2\pi)^3\delta^3(\vec{k}-\vec{q}_{1,n})F_n(\vec{q}_1,\vec{q}_2,...\vec{q}_n)\delta^{(1)}_{\vec{q}_1}\delta^{(1)}_{\vec{q}_2}...\delta^{(1)}_{\vec{q}_n}\;,\nonumber\\
    \theta^{(n)}(\vec{k})=\int_{\vec{q}_1...\vec{q}_n}(2\pi)^3\delta^3(\vec{k}-\vec{q}_{1,n})G_n(\vec{q}_1,\vec{q}_2,...\vec{q}_n)\delta^{(1)}_{\vec{q}_1}\delta^{(1)}_{\vec{q}_2}...\delta^{(1)}_{\vec{q}_n}\;,
\end{align}
where $\vec{q}_{1,n}=\sum_{i=1}^n\vec{q}_i$ and $F_n, G_n$ are kernel derived recursively from Eq.\eqref{eq:fluid_eq} order by order.  Formally, SPT might be expected to work as long as the perturbations $\delta \lesssim \mathcal O(1)$. The scale, or wavenumber at which typical fluctuations are $\mathcal O(1)$ is referred to as the non-linear scale and denoted by $k_{\rm NL}$. However, by comparing to simulations, it has been shown that the SPT treatment breaks down for wavenumbers well below $k_{\rm NL}$. This led to the development of EFTofLSS which correctly incorporates the effective treatment of short-scale physics and matches well with data. However, for what follows for the next couple of sections, we do not need to consider EFTofLSS for the evolution of Dark Matter. The SPT kernels introduced here will be sufficient for the analysis and we will discuss the modifications of the Dark Matter equations in the context of EFTofLSS in Sec.\ref{ssec:EFT}.

\subsection{Time local Bias Expansion}\label{ssec:loc_bias_exp}
Baryonic matter, such as galaxies, traces the underlying dark matter field as a result of which the number density of these "tracers" is correlated with the dark matter overdensity. Due to the absence of an equation of motion for tracers, there is no first principle way to get the tracer overdensity. However, we know that the tracer overdensity can be affected by dark matter fields only locally and therefore, can only depend on the locally measurable quantities constructed out of the dark matter fields. Due to rotational invariance and the equivalence principle, the locally measurable quantities are the tidal field ($\partial_i\partial_j\Phi$) and velocity gradient ($\partial_iv_j$). Tracer overdensity is a scalar quantity, therefore it can depend only on scalars constructed out of the tidal field and velocity gradient.

This allows us to write the tracer overdensity as follows\,\cite{Senatore:2014eva,McDonald:2009dh},
\begin{align}\label{eq:f_exp}
    \delta_g(\xvec,t)=b_{\partial^2\Phi}(t)\partial^2\Phi(\xvec,t)+b_{\partial_iv_i}(t)\partial_iv_i(\xvec,t)+b_{\partial_i\partial_j\Phi\partial_iv_j}(t)\partial_i\partial_j\Phi\partial_iv_j(\xvec,t)+...\;,
\end{align}
where $\delta_g\equiv \frac{n_g-\Bar{n}_g}{\Bar{n}_g}$, $n_g$ being the number density of galaxies and $b$'s are time-dependent coefficients also called "bias" coefficients which can be measured from simulation or galaxy data by galaxy-galaxy correlation or galaxy-dark matter cross correlations. We have just written the leading order terms in Eq.\eqref{eq:f_exp} and not the terms containing higher derivatives which are also present\footnote{In this work we will not be considering higher derivative and stochastic terms in the bias expansion for tracers.}. Note that Eq.\eqref{eq:f_exp} is not the most general form for tracer overdensity since it is local-in-time. The most general form for tracer overdensity is time non-local and will be discussed later in Sec.\,\ref{sec:nloc_bias_exp}.
We can write Eq.\eqref{eq:f_exp} in a more familiar and compact form as,
\begin{align}\label{eq:tracer_loc_exp}
 \delta_g(\vec{x},t)=\sum_{\mathcal{O}_a}b_{\mathcal{O}_a}(t)\mathcal{O}_a(\vec{x},t)\;,
\end{align}
where $b_{\mathcal{O}_a}(t)$ are time-dependent bias coefficients and $\mathcal{O}_a$ are scalar operators made of long wavelength dark matter fields that are allowed by the symmetry i.e. $SO(3)$ invariance and the equivalence principle. Therefore, the operators $\mathcal{O}_a$ are $SO(3)$ scalars constructed from dimensionless tidal field and velocity gradients defined as,
\begin{align}\label{eq:tid_vel_grad}
  r_{ij}\equiv\frac{2\partial_i\partial_j\Phi}{3\Omega_m a^2H^2}\hspace{2cm}p_{ij}\equiv-\frac{D}{a\dot{D}}\partial_iv_j\;,
\end{align}
where $\Omega_m$, $a$, and $H$ are matter fraction, scale factor, and Hubble parameter respectively. $D$ is the linear growth factor defined below Eq.\eqref{eq:ansatz_series}. We can construct $SO(3)$ invariant operators by taking products of the operators given in Eq.\eqref{eq:tid_vel_grad} contracting all indices to make an $SO(3)$ scalar as,
\begin{align}\label{eq:SO3_def}
\delta\equiv\delta_{ij}^Kr_{ij},\hspace{0.5cm}\theta\equiv\delta_{ij}^Kp_{ij},\hspace{0.5cm}r^2\equiv r_{ij}r_{ji}\;.
\end{align}
At each order in dark matter overdensity, the allowed set of operators are all possible $SO(3)$ scalars constructed from $r_{ij}$ and $p_{ij}$ in the manner shown in Eq.\eqref{eq:SO3_def} which are relevant at that order. Both the tidal field and velocity gradient start at linear order, so an operator such as $r^2$ or $rp$ will start at second order in dark matter overdensity and hence will be relevant at second order and not at lower orders. This implies that, at each order, there will be a finite number of operators appearing in the bias expansion. 

An operator $\mathcal{O}_a$ can contribute to any given order $n$ if it starts at an order greater than or equal to $n$. For example, the operator $rp$ can be evaluated at order three as,
\begin{align}\label{eq:sample_kernel}
    (rp)^{(3)}=r^{(1)}p^{(2)}+r^{(2)}p^{(1)}\;,
\end{align}
where each operator is evaluated at the same point $\xvec$. We can express Eq.\eqref{eq:sample_kernel} in momentum space and use Eq.\eqref{eq:tid_vel_grad} and Eq.\eqref{eq:kernels} to write the kernel for $(rp)^{(3)}$. Similarly, in momentum space, any operator $\mathcal{O}_a$ that occurs in Eq.\eqref{eq:tracer_loc_exp} at $n^{th}$ order can be written as convolution over linear dark matter fields with a kernel $K^{\mathcal{O},a}_n$ as,\footnote{Let us note that to compute the kernels for operators we have used \eqref{eq:kernels} which are the solutions of SPT.}
\begin{align}\label{eq:On_def}
\mathcal{O}_a^n(\vec{k})=\int_{\vec{q}_1...\vec{q}_n}(2\pi)^3\delta^3(\vec{k}-\vec{q}_{1,n})K^{\mathcal{O},a}_n(\vec{q}_1,\vec{q}_2,...\vec{q}_n)\delta^{(1)}_{\vec{q}_1}\delta^{(1)}_{\vec{q}_2}...\delta^{(1)}_{\vec{q}_n}\;,
\end{align}
where $\vec{q}_{1,n}=\sum_{i=1}^n\vec{q}_i$ and $K^{\mathcal{O},a}_n(\vec{q}_1,\vec{q}_2,...\vec{q}_n)$ are rational functions of the momenta $\vec{q}_i$. So essentially, in momentum space the difference in the operators appearing in the bias expansion as given in Eq.\eqref{eq:tracer_loc_exp}, comes through their kernels. However, it can be shown that not all operators at a given order are independent. For example at the second order, we have the following set of operators that contribute to the bias expansion,
\begin{align}
    \{\delta^{(2)},\theta^{(2)},(\delta^2)^{(2)},(\delta\theta)^{(2)},(\theta^2)^{(2)},(r^2)^{(2)},(rp)^{(2)},(p^2)^{(2)}\}\;.
\end{align}
However, in momentum space, there exists a linear relation satisfied by the following four operators,
\begin{align}\label{eq:lin_rel_order_2}
    \delta^{(2)}-\theta^{(2)}-\frac{2}{7}(\delta^2-r^2)^{(2)}=0\;,
\end{align}
which implies that one of these operators is degenerate and its kernel can be written in terms of the other three. Hence, it should not appear in the bias expansion given in Eq.\eqref{eq:tracer_loc_exp} with an independent bias coefficient. In this way, we can remove all the degenerate operators from the bias expansion if a linear relation exists among subsets of the operators. This will lead us to an independent set of operators that constitutes the "basis" of the operator space at any given order.


\subsection{Selection operators}\label{ssec:sel_op}

In this section we introduce some new $SO(2)$ invariant operators, which we call selection operators, in the bias expansion as given in Eq.\eqref{eq:tracer_loc_exp}. We will explain the mathematical structure for these operators and discuss physical situations in which they may naturally arise.

We will show below that selection operators have distinct structures compared to the usual $SO(3)$ operators considered in Eq.\eqref{eq:tracer_loc_exp}. This implies that selection operators appear with independent coefficients in the bias expansion of tracers. So now the bias expansion takes the following form,
\begin{align}\label{eq:bias_exp_selec}
\delta_g(\vec{x},t)=\sum_{a}b_{\mathcal{O}_a}(t)\mathcal{O}_a(\vec{x},t)+\sum_{m}c_{\mathcal{P}_m}(t)\mathcal{P}_m(\vec{x},t)\;,
\end{align}
where $c_{\mathcal{P}_m}$ are new bias coefficients and $\mathcal{P}_m(\vec{x},t)$ are new operators that are invariant under rotation about a vector, which is generally taken to be the direction of line-of-sight (LoS). Due to the presence of a LoS, the symmetry is reduced to $SO(2)$ invariance along the LoS. $SO(2)$ invariance and the equivalence principle ensures that $\partial_i\partial_j\Phi$ and $\partial_iv_j$ and their projections along the line of sight are also locally observable quantities. Therefore, selection operators are constructed from the same building blocks as the $SO(3)$ operators i.e. the tidal field and the velocity gradient as given in Eq.\eqref{eq:tid_vel_grad}. From these we construct two indexed objects which are then contracted with unit vectors ($\hat{z}$) to make $SO(2)$ scalars such as,
\begin{align}\label{eq:sel_op}
    \delta_z\equiv r_{ij}\hat{z}^i\hat{z}^j,\hspace{0.5cm} \eta\equiv p_{ij}\hat{z}^i\hat{z}^j,\hspace{0.5cm}(rp)_\parallel\equiv r_{ij}p_{jk}\hat{z}^i\hat{z}^k\;,
\end{align}
where $\hat{z}$ is the unit vector along the line of sight. The operators that appear in bias expansion will consist of all such operators as given in Eq.\eqref{eq:sel_op} and their products. 

Note that the $SO(2)$ scalars, being a projection of $SO(3)$ tensors allow for more operators to be independent of each other which enhances the operator space. For example, consider $r_{ij}$ and $\delta^K_{ij}\delta$ which are same operator, $\delta$ if we try to construct $SO(3)$ scalars by taking trace. However, if we construct $SO(2)$ scalars by taking projections as $r_{ij}\hat{z}^i\hat{z}^j$ and $(\delta_{ij}^K\delta)\hat{z}^i\hat{z}^j$, then they are independent operators and occur with independent bias coefficients as given in Eq.\eqref{eq:bias_exp_selec}. Due to this fact, one might expect the time non-locality to appear at lower order. We have found that this is indeed the case and we have discussed this in detail in Sec.\,\ref{sec:nloc_sel_op}.

Physically, these selection operators are crucial while comparing the theoretical predictions with observations. This is due to the fact that we observe galaxies through the light that reaches us. The probability that we will observe a photon emitted from a galaxy in a particular direction depends on the line-of-sight of observation. These effects can be incorporated in the bias expansion by including new line of sight dependent operators\,\cite{Desjacques:2018pfv} which are precisely the selection operators given in Eq.\eqref{eq:bias_exp_selec}. In the context of galaxies, a subset of these selection operators appear when we move to galaxy clustering in redshift space. 

However, in other contexts such as when writing an EFT for the flux field from Lyman-$\alpha$ forest, we need to include all possible selection operators\,\cite{Ivanov:2023yla} in the bias expansion of the flux fluctuations. This happens due to the non-linear mapping between the flux field ($F$) and the optical depth ($\tau$), which is a biased tracer of the underlying dark matter field. The non-linear mapping is given by an exponential as $F \propto e^{-\tau}$. Owing to this exponential map, a "renormalised" EFT of flux overdensity field, $\delta_F=(F-\Bar{F})/\Bar{F}$, with the optical depth taken to be in redshift space, contains all possible selection operators. We have commented on this in a bit more detail in Appendix.\,\ref{ap:renorm}. This way of generating all possible selection operators is termed as the "top-down" approach to writing EFTs as opposed to the "bottom-up" approach that we have discussed so far. 

\subsection{$\Pi$ basis}\label{sssec:Pi_basis}

In this section, we describe an alternate system of writing operators constructed out of long wavelength dark matter fields, the $\Pi$ system \cite{Mirbabayi:2014zca}. However, the effectiveness of $\Pi$ system will be shown in capturing the time non-locality in biased tracers or dark matter as discussed later in Secs.\,\ref{sec:nloc_bias_exp} and \ref{sec:nloc_dm} respectively. The usual method to compute the set of independent non-local operators for biased tracers is discussed in\,\cite{DAmico:2022ukl,Donath:2023sav}. This method is very tedious and contains a lot of redundancies. However, we point out that when working with the $\Pi$ system\,\cite{Mirbabayi:2014zca}, the calculation of determining the set of independent non-local operators is much more direct and efficient.

For biased tracers of the dark matter field, such as halos and galaxies, \cite{Mirbabayi:2014zca} proposed a system for writing the relevant operators which stems from the fact that convective derivative of a locally observable operator is also locally observable\,\cite{Mirbabayi:2014zca,Desjacques:2016bnm}. We will use this system to write down the EFT operators appearing in the effective stress tensor relevant for the dark matter field itself and also to write the operators appearing in the bias expansion for tracers. The reason to do this is that they have a nice connection with time non-locality which will be discussed in later sections. In this system we define
\begin{equation}\label{pi1}
\Pi_{ij}^{[1]}\equiv r_{ij}.
\end{equation} We also define
\cite{Mirbabayi:2014zca}, 
\begin{align}\label{eq:Pi_n}
    \Pi_{ij}^{[n]}=\frac{1}{(n-1)!}\Bigg[(\mathcal{H}f)^{-1}\frac{D}{D\eta}\Pi_{ij}^{[n-1]}-(n-1)\Pi_{ij}^{[n-1]}\Bigg]\;,
\end{align}
where $f\equiv d\ln D/d\ln a$ and $\mathcal{H}$ are the logarithmic growth factor and Hubble parameter (expressed in conformal time) respectively and convective derivative is defined as
\begin{align}\label{eq:con_der_def}
   \frac{D}{D\eta}\equiv\frac{\partial}{\partial\eta}+v^i\partial_i\;.
\end{align}
Here the basic idea is,  a convective derivative of a locally measurable quantity is also locally measurable so we can act with convective derivative on the tidal field and get another locally measurable operator\footnote{ Locally measurable quantities are tidal field ($\partial_i\partial_j\Phi$) and velocity gradient ($\partial_iv_j$) as a uniform acceleration given by $\partial_i\Phi$ is not locally observable due to the equivalence principle.}. So in this system, instead of tidal field and velocity gradient as given in Eq.\eqref{eq:tid_vel_grad}, the building blocks for constructing $SO(3)$ scalars are $\Pi_{ij}^{[n]}$ operators. Using \eqref{eq:Pi_n}, we can construct operators by taking trace of $\Pi_{ij}^{[n]}$ as Tr$[\Pi_{ij}^{[1]}]=\Pi_{ii}^{[1]}$, or other two indexed objects made from $\Pi^{[n]}$ as, Tr$[\Pi^{[2]}\Pi^{[1]}]\equiv$Tr$[\Pi_{ij}^{[2]}\Pi_{jk}^{[1]}]=\Pi_{ij}^{[2]}\Pi_{ji}^{[1]}$ and products of the traces of individual $\Pi_{ij}^{[n]}$ appropriately.







\section{Non-locality in dark matter}\label{sec:nloc_dm}
SPT has been sucessful in describing LSS at very large scales but as we approach shorter distances i.e. higher wavenumber, linear SPT starts to diverge from the results of fully nonlinear $N$-body simulations. Even for $k<k_{\rm NL}$, i.e. quasilinear scales where $\delta<\mathcal O(1)$, higher loop calculations of the SPT power spectrum do not converge to the correct result\,\cite{Carrasco:2013mua}. 
To cure this, the correct treatment is to smooth out the equations of motion over a length scale above the non-linear scale. The effect of this smoothing out procedure is to include ``effective" contributions from non-linear scale which is the subject of EFTofLSS \cite{Carrasco:2012cv}. In this section, we will discuss the governing equations for EFTofLSS and the effective stress tensor which comes as result of defining the EFT. We will also discuss  significance of this effetcive stress tensor in the context of  time non-locality.

\subsection{EFT of LSS}\label{ssec:EFT}
In this sub section, we continue the discussion outlined in Sec.\,\ref{ssec:SPT}. In order to define an EFT pertaining to the equations of motion given in Eq.\eqref{eq:fluid_eq}, we need to integrate the short scale modes. After integrating out the short modes in the fluid equations, we get,
\begin{align}\label{eq:fluid_eq_long}
    \delta_l'(\xvec,\eta) &+ \nabla\cdot[\vec{v}_l(\xvec,\eta)(\delta_l(\xvec,\eta) + 1)] = 0\;,\nonumber\\
    v_{i,l}'(\xvec,\eta) + &\mathcal{H} v_{i,l}(\xvec,\eta) + (\vec{v}_l(\xvec,\eta).\nabla)v_{i,l}(\xvec,\eta) = -\nabla_i\Phi_l(\xvec,\eta)-\frac{1}{\rho_l}\partial_j\tau_{ij}(\xvec,t)\;,
\end{align}
where the subscript "$l$" indicates the long wavelength quantities and $\tau_{ij}$ is an effective stress tensor generated from the smoothing and contains all the information about short scales. The stress tensor regulates the loops in SPT which access the non-linear scales through integrals in Eq.\eqref{eq:kernels} and keeps the theory finite. One can write the effective stress tensor in terms of the long wavelength dark matter fields as follows,
\begin{align}\label{eq:tau_loc_exp}
\tau_{ij}(\xvec,t)=\sum_{\mathcal{O}^m_{ij}}c_{\mathcal{O}^m_{ij}}(t)\mathcal{O}^m_{ij}(\xvec,t)\;,
\end{align}
where $c_{\mathcal{O}^m_{ij}}(t)$ are EFT coefficients that are fixed by data or simulations and $\mathcal{O}^m_{ij}$ are operators constructed from locally observable long wavelength dark matter fields. Again for the case of dark matter, locally observable fields are tidal fields ($\partial_i\partial_j\Phi$) and the velocity gradient ($\partial_iv^j$). So the operators appearing in Eq.\eqref{eq:tau_loc_exp} are constructed by contracting these long wavelength local observable. For example, at lowest order 
\begin{align}
\tau_{ij}^{(1)}=c_1(t)\partial_i\partial_j\Phi^{(1)}+c_2(t)\delta^K_{ij}\partial_k\partial^k\Phi^{(1)}\;,
\end{align}
where $c_i(t)$ are EFT coefficients and $\delta_{ij}^K$ is the Kronecker delta function respectively. We can see from Eq.\eqref{eq:tau_loc_exp}, that it is local in time as both r.h.s and l.h.s are evaluated at the same time. Note that Eq.\eqref{eq:tau_loc_exp} is not the most general expression for the effective stress tensor. The stress tensor is actually non-local in time for reasons we describe below. Therefore, the most general expression for the effective stress tensor is also non-local in time and is given later in Eq.\eqref{eq:tau_nloc_exp}. To write down a local effective theory for the long mode, it is important to note that there should be a separation of length scale. However, for dark matter, even though there is a separation of length scale in the spatial domain but there is no separation of scale in temporal domain. This leads to time non-locality in the effective description of the dark matter after integrating out the short modes.
    To understand the absence of scale separation 
    in the time domain, we can also look at the linear form of Eq.\eqref{eq:fluid_eq}. If we linearise the equations given in Eq.\eqref{eq:fluid_eq} and substitute the Euler equation in the continuity equation, we get the following linear growth equation for $\delta$,
\begin{align}\label{eq:lin_eq_del}
    \delta^{\prime\prime}(\vec{k},\eta) + \mathcal{H}(\eta)\delta^\prime(\vec{k}, \eta) - \frac{3}{2}\Omega_m(\eta)\mathcal{H}^2(\eta)\delta(\vec{k}, \eta) = 0\;,
\end{align}
where $^\prime$ denotes a derivative w.r.t the conformal time. From Eq.\eqref{eq:lin_eq_del}, we can see that there is no explicit $\vec{k}$ dependence and hence both the long and short modes follow the same dynamics. This is another manifestation of the fact that there is no scale separation in the time domain even though there is a scale separation in the momentum domain through $k_\text{NL}$\footnote{ 
This can also be seen from the discussion above Eq.\eqref{eq:fluid_eq}, where the dark matter is modelled as a collisionless (pressureless) fluid. Due to the absence of pressure, the speed of sound vanishes in such a fluid. Hence, there does not exist any dispersion relation which relates the wavenumber ($\vec{k}$) with the frequency. However, after integrating out the short modes we get an effective stress tensor on the r.h.s of Eq.\eqref{eq:lin_eq_del}. This introduces an effective pressure and the wavenumber and frequency are related through a dispersion relation.}. Therefore, as mentioned earlier, when we smooth out the short modes to write an EFT, the physics is local in space domain but non-locality is introduced in the time domain. The time non-locality in the context of EFTofLSS for dark matter was first pointed in\,\cite{Carrasco:2013mua}. The time non-locality is incorporated in the effective stress tensor which can be written as\,\cite{Carrasco:2013mua,DAmico:2022ukl},
\begin{align}\label{eq:tau_nloc_exp}
\tau_{ij} (x,t)=\sum_{\mathcal{O}^m_{ij}}\int^tdt'c_{\mathcal{O}^m_{ij}}(t,t')\mathcal{O}^m_{ij}(\vec{x}_{\text{fl}}(\vec{x},t,t'),t')\;,
\end{align}
where $c_{\mathcal{O}^m_{ij}}(t,t')$ are time dependent kernels and $\vec{x}_{\text{fl}}$ is defined implicitly through the following relation,
\begin{align}\label{eq:xfl_exp_1}
    \vec{x}_{\text{fl}}(\vec{x},t,t')=\vec{x}-\int_{t'}^t \frac{dt''}{a(t'')}\vec{v}(t'',\vec{x}_{\text{fl}}(\vec{x},t,t''))\;.
\end{align}
 where $\vec{x}_{\text{fl}}$ is the coordinate labelling the past trajectory of the fluid element that ended up at position $\xvec$ at time $t$ and $\vec{v}$ is the velocity of the fluid element along the past trajectory. Note that Eq.\eqref{eq:tau_nloc_exp} reduces to expansion given in Eq.\eqref{eq:tau_loc_exp} when $c_{\mathcal{O}^m_{ij}}(t,t')=c_{\mathcal{O}^m_{ij}}(t)\delta(t-t')$.

As we can see from Eq.\eqref{eq:tau_nloc_exp}, that due to the time integration, Eq.\eqref{eq:tau_nloc_exp} is non-local in time as compared to Eq.\eqref{eq:tau_loc_exp} which is local in time. However, we can use Eq.\eqref{eq:xfl_exp_1} iteratively in Eq.\eqref{eq:tau_nloc_exp} and Taylor expand it around $\xvec$ to get $\mathcal{O}_i$ up to any arbitrary order. After expanding about $\xvec$, we will still be left with the integral given in Eq.\eqref{eq:xfl_exp_1}. Assuming no vorticity i.e. $v^i \propto \partial^i\theta$, we can perform the integral given in Eq.\eqref{eq:xfl_exp_1} and get an expression arranged in powers of $\frac{D(t')}{D(t)}$\,\cite{DAmico:2022ukl,Donath:2023sav}. This process removes the time non-locality but introduces space non-locality and the structure of operators appearing in the expansion given in Eq.\eqref{eq:tau_nloc_exp} changes\,\cite{DAmico:2022ukl}.

Restricting ourselves upto order $n$ in dark matter overdensity, we get\,\cite{Donath:2023sav,DAmico:2022ukl},
\begin{align}\label{eq:Oin_exp_dm}
     [ \mathcal{O}^m_{ij} ( \vec{x}_{\text{fl}} ( \vec{x} , t , t' ) , t')]^{(n)}  =   \sum_{\alpha = 1}^{n - m +1} \left( \frac{D(t')}{D(t)} \right)^{\alpha+m-1} \mathbb{C}_{\mathcal{O}^m_{ij}, \alpha}^{(n)} (\vec{x} , t )  \ .
\end{align}
where $\mathbb{C}_{\mathcal{O}^m_{ij}, \alpha}^{(n)}$ are the new operators stemming from each operator $\mathcal{O}^m_{ij}$ upon the Taylor expansion along fluid trajectory. Substituting the form of $(\mathcal{O}^m_{ij})^{(n)}$ from Eq.\eqref{eq:Oin_exp_dm} into Eq.\eqref{eq:tau_nloc_exp} we get the following expansion that is local in time,
\begin{align}\label{eq:tau_nloc_loc_exp}
    \tau_{ij} ( \xvec , t) = \sum_{\mathcal{O}_i} \sum_{\alpha = 1}^{n-m+1} c_{\mathcal{O}^m_{ij},\alpha}(t) \,\mathbb{C}^{(n)}_{\mathcal{O}^m_{ij},\alpha} ( \xvec , t)\;,
\end{align}
where, 
\begin{align}\label{eq:tau_nloc_bias_coeff}
    c_{\mathcal{O}^m_{ij},\alpha} ( t ) \equiv \int^t dt' c_{\mathcal{O}^m_{ij}} ( t , t' ) \left( \frac{D(t')}{D(t)} \right)^{\alpha+m-1 } \;,
\end{align}
are the time dependent bias coefficients similar to what is given in the local expansion in Eq.\eqref{eq:tau_loc_exp}. 

Note that Taylor expanding Eq.\eqref{eq:tau_nloc_exp} using Eq.\eqref{eq:xfl_exp_1} introduces the convective derivative structure similar to what is given in Eq.\eqref{eq:con_der_def}. Therefore, one expects that the basis obtained in this manner is generically equivalent to the $\Pi$ basis discussed in Sec.\,\ref{sssec:Pi_basis}, which is constructed by acting convective derivatives on local operators. We will see later that indeed both systems produce equivalent bases at each order.
\begin{table}[h!]
       \center
       \renewcommand{\arraystretch}{1.25}
       \begin{tabular}{|c|c|c|c| } 
           \hline 
            ~~$ $&$ $&$ $&$ $\\
           ~~$ $&$\textbf{Local Basis} $&$ \textbf{Non-local basis}$&$\Pi$-\textbf{basis} \\
            ~~$ $&$ $&$ $&$ $\\
        \hline 

  ~~$ $&$ $&$ $&$ $\\
       
           ~~$ 1^{st}$ order &$r_{ij}, \delta_{ij}^K\delta $&$\Cops{1}{r_{ij},1},\Cops{1}{\delta_{ij}^K\delta,1} $&$\Pi_{ij}^{[1]},\delta_{ij}^K\text{Tr}[\Pi^{[1]}] $\\
            ~~$ $&$ $(\textbf{2})&$ $(\textbf{2})&$ $(\textbf{2})\\

            \hline 
             ~~$ $&$ $&$ $&$ $\\
           ~~$2^{nd} $ order &$r_{ij},p_{ij},r_{ij}\delta,(r^2)_{ij} $&$\Cops{2}{p_{ij},1},\Cops{2}{p_{ij},2},\Cops{2}{r_{ij},2},\Cops{2}{p_{ij}\delta,2} $&$\Pi_{ij}^{[1]},\Pi_{ij}^{[1]}\text{Tr}[\Pi^{[1]}],\Pi_{ik}^{[1]}\Pi_{kj}^{[1]},\Pi_{ij}^{[2]} $\\
            ~~$ $&$ $&$ $&$ $\\
           
            ~~$ $&$ \delta_{ij}^K(\delta, \delta^2,r^2) $&$\delta_{ij}^K(\Cops{2}{\delta,1},\Cops{2}{\delta,2},\Cops{2}{\theta,2}) $&$\delta_{ij}^K(\text{Tr}[\Pi^{[1]}],\text{Tr}[\Pi^{[1]}]^2, $\\
          
         ~~$ $&$ $&$ $&$\text{Tr}[\Pi^{[1]}\Pi^{[1]}]) $\\
 ~~$ $&$ $(\textbf{7})&$ $(\textbf{7})&$ $(\textbf{7})\\

            \hline 
             ~~$ $&$ $&$ $&$ $\\
           ~~$3^{rd} $ order&$r_{ij},p_{ij},p_{ij}\delta,p_{ij}\theta,r_{ij}\delta $&$\Cops{3}{r_{ij},1},\Cops{3}{r_{ij},2},\Cops{3}{r_{ij},3},\Cops{3}{p_{ij},2} $&$\Pi_{ij}^{[1]},\text{Tr}[\Pi^{[1]}]\Pi_{ij}^{[1]},\Pi_{ik}^{[1]}\Pi_{kj}^{[1]},\Pi_{ij}^{[2]}, $\\
            ~~$ $&$ $&$ $&$ $\\

           ~~$ $&$(r^2)_{ij},(p^2)_{ij},p_{ij}\delta^2,(r^3)_{ij}, $&$\Cops{3}{p_{ij},3},\Cops{3}{p_{ij}\delta,1},\Cops{3}{p_{ij}\delta,2},\Cops{3}{p_{ij}\theta,2}, $&$\Pi_{ij}^{[1]}\text{Tr}[\Pi^{[1]}]^2,[\Pi^{[1]}\Pi^{[1]}\Pi^{[1]}]_{ij}, $\\
           
            ~~$ $&$ $&$ $&$ $\\

             ~~$ $&$ $&$ $&$ $\\

           ~~$ $&$\delta_{ij}^K(\delta,\delta^2,\delta\theta,r^2,rp,\delta^3,r^3) $&$\Cops{3}{r_{ij}\delta,2},\Cops{3}{(r^2)_{ij},2}, $&$[\Pi^{[1]}\Pi^{[1]}]_{ij}\text{Tr}[\Pi^{[1]}],\text{Tr}[\Pi^{[1]}]\Pi_{ij}^{[2]}, $\\

            ~~$ $&$ $&$ $&$ $\\

             ~~$ $&$ $(\textbf{16})&$\delta_{ij}^K(\Cops{3}{\delta,1},\Cops{3}{\delta,2},\Cops{3}{\delta,3} $&$[\Pi^{[1]}\Pi^{[2]}]_{ij},\Pi_{ij}^{[3]},\delta_{ij}^K(\text{Tr}[\Pi^{[1]}], $\\

              ~~$ $&$ $&$ $&$ $\\

              ~~$ $&$ $&$\Cops{3}{\theta,2},\Cops{3}{\theta,3},\Cops{3}{\delta^2,2},\Cops{3}{r^2,2}) $&$\text{Tr}[\Pi^{[1]}]^2,\text{Tr}[\Pi^{[1]}\Pi^{[1]}],\text{Tr}[\Pi^{[2]}], $\\

               ~~$ $&$ $&$ $&$ $\\

               ~~$ $&$ $&$ $(\textbf{17})&$\text{Tr}[\Pi^{[1]}]^3,\text{Tr}[\Pi^{[1]}\Pi^{[1]}\Pi^{[1]}], $\\

                ~~$ $&$ $&$ $&$\text{Tr}[\Pi^{[1]}\Pi^{[1]}]\text{Tr}[\Pi^{[1]}]) $\\

                ~~$ $&$ $&$ $&$ $(\textbf{17})\\

                 ~~$ $&$ $&$ $&$ $\\
          
        \hline 

       \end{tabular}
     \caption{List of the local, non-local and $\Pi$ basis for the effective stress tensor is shown. We can see that the local and non-local basis differ from each other at third order. In writing local operators, we have used the notation given in Eq.\eqref{eq:tid_vel_grad} and \eqref{eq:SO3_def}. The notation for writing non-local operators has been taken from\,\cite{Donath:2023sav, DAmico:2022ukl}, and is explained in Eqs.\eqref{eq:tau_nloc_exp} and \eqref{eq:Oin_exp_dm}.}
       \label{tab:dm_nloc}
   \end{table}

Time non-locality enters the dynamics of long wavelength dark matter fields through the stress tensor in Eq.\eqref{eq:fluid_eq_long}. The effective stress tensor has the local in time expansion as given in Eq.\eqref{eq:tau_loc_exp} and the non-local in time expansion in Eq.\eqref{eq:tau_nloc_exp}. At each order, only a finite number of independent operators appear in the expansion. These form the basis of operators at each order. We have found that up to second order, the time local and time non-local stress tensor as given in Eq.\eqref{eq:tau_loc_exp} and Eq.\eqref{eq:tau_nloc_exp} respectively, are equivalent. However, they differ from each other at third order which we discuss in the next section. We have analysed the structure of $\partial_j\tau_{ij}$ as well, that appears in Eq.\eqref{eq:fluid_eq_long}, and which we refer to as vector structure. We have found that time non-locality appears in the vector structure at third order as well. We have discussed this in bit more detail in Appendix.\,\ref{ap:vec_op}.
 
\subsection{Time non-locality in stress tensor ($\tau_{ij}$)}\label{subsec:tensor_nloc}
In this section, we discuss the disparity between the time local and time non-local basis of operators that appear in the effective stress tensor. We have found that the time local basis differs from the time non-local basis at third order. This result is summarised in Table.\,\ref{tab:dm_nloc} where we have listed the independent set of operators in the time local and time non-local basis up to third order.  

We can see from Table.\,\ref{tab:dm_nloc}, that up to second order the time local and non-local basis are equivalent to each other. However, at third order, the local basis consists of 16 operators while the non-local basis contains 17 independent operators. Hence, we find time non-locality in the effective stress tensor at third order. As a crosscheck of non-local structures, we have computed the independent set of operators in the $\Pi$ system as well. Note that the general structure of effective stress tensor has been obtained up to third order in terms of $\Pi$ basis elsewhere\,\cite{Vlah:2019byq}\footnote{We thank the referee for pointing this out to us.}. We have found that the $\Pi$ basis also contains 17 independent operators at third order which is same as the non-local basis. Our result for $\Pi$ basis agrees with that obtained in\,\cite{Vlah:2019byq}\footnote{The list of operators given in\,\cite{Vlah:2019byq} contains 11 tensor structures. However, it can be shown that one of the operators can be written as a linear combination of other tensor structures and $\delta_{ij}$ times trace operators. This makes the list of independent tensor structures at third order to be 10 which is given in Table.\ref{tab:dm_nloc}.}. We have checked that both the non-local and $\Pi$ bases are related by an invertible map. The list of independent $\Pi$ operators is also given in Table.\,\ref{tab:dm_nloc}.\footnote{It's worth mentioning that despite our focused examination on the non-locality in stress tensor, a similar rationale can be applied to any generic second-rank tensor that the non-locality will shift to third order.} Let us note that, at third order, $(\Pi^{[2]}_{ij})^{(3)}$ which can be treated as a non-local operator (due to the invertible map), can't be written in terms of the local operators listed in Table.\,\ref{tab:dm_nloc}\footnote{We note that our list of operators included in the trace part of $\tau_{ij}$ contains $\text{Tr}[\Pi^{[2]}]$ whereas the corresponding list in \cite{Vlah:2019byq} features $\text{Tr}[\Pi^{[2]}\Pi^{[1]}]$. We would like to emphasise that both are equivalent choices at third order.}.

The dynamical equations, however, does not contain $\tau_{ij}$ directly but it's first derivative in the form $\partial_j\tau_{ij}$ which we have called as vector structure. We have analysed the structure of local as well as non-local expansion of $\partial_j\tau_{ij}$. We have found that non-locality shows up in vector structure also at third order. Please see Appendix.\,\ref{ap:vec_op} for details.


\section{Non-locality in biased tracers in real space}\label{sec:nloc_bias_exp}
Having discussed the local bias expansion in Sec.\,\ref{ssec:loc_bias_exp}, we can discuss time non-locality in biased tracers. In this section, we will layout the physics of biased tracers and discuss the origin of time non-locality among biased tracers. We will mainly be reviewing ideas presented in\,\cite{Senatore:2014eva,DAmico:2022ukl,Donath:2023sav}. We also discuss that writing operators in the $\Pi$ system, which is described in Sec.\,\ref{sssec:Pi_basis}, makes the task of finding the non-local basis much more efficient.

As mentioned earlier, tracers such as galaxies evolve over the background of dark matter fields. However, it has been shown that the time of virialisation for galaxies or halos is around one-tenth of the time of evolution of dark matter fields i.e. the time taken by a long mode to reach the non-linear scale\,\cite{Senatore:2014eva}. This implies that galaxy formation does not occur instantaneously but takes a small but finite time period\footnote{If galaxies are formed instantaneously, the background dark matter fields do not change significantly during the formation time. Due to this, the galaxy distribution and hence, the number density will be affected by dark matter fields at the same time. For such a scenario, the bias expansion given in Eq.\eqref{eq:tracer_loc_exp} is appropriate as it is local in time in the sense that both r.h.s and l.h.s is evaluated at the same time.}. Since, galaxy formation is not instantaneous, the tracers are affected by the dark matter fields not only at the same time but also through the whole past history along the fluid trajectory.  Therefore, it has been argued\,\cite{Senatore:2014eva}, that the actual bias expansion should be non-local in time and the r.h.s of Eq.\eqref{eq:tracer_loc_exp} should be generalized to an integral over the fluid trajectory of dark matter as follows,    
\begin{align}\label{eq:tracer_nloc_exp}
\delta_g(\vec{x},t)=\sum_{\mathcal{O}_i}\int^{t}dt'c_{\mathcal{O}_i}(t,t')\mathcal{O}_i(\vec{x}_{\text{fl}}(\vec{x},t,t'),t')\;,
\end{align}
where $c_{\mathcal{O}_i}(t,t')$ are kernels having support over one Hubble time and $\vec{x}_{\text{fl}}$ is defined in Eq.\eqref{eq:xfl_exp_1}. Note that taking $c_{\mathcal{O}_i}(t,t')=c_{\mathcal{O}_i}(t)\delta(t-t')$ gives us the local-in-time bias expansion given in Eq.\eqref{eq:tracer_loc_exp}.

Following the same procedure as we did in Sec.\ref{ssec:EFT}, we can use Eq.\eqref{eq:xfl_exp_1} iteratively to Taylor expand the operator $\mathcal{O}_i$ in Eq.\eqref{eq:tracer_nloc_exp}. Restricting ourselves upto order $n$ in dark matter overdensity, we get\,\cite{Donath:2023sav,DAmico:2022ukl},
\begin{align}\label{eq:Oin_exp}
     [ \mathcal{O}_i ( \vec{x}_{\text{fl}} ( \vec{x} , t , t' ) , t')]^{(n)}  =   \sum_{\alpha = 1}^{n - i +1} \left( \frac{D(t')}{D(t)} \right)^{\alpha+i-1} \mathbb{C}_{\mathcal{O}_i, \alpha}^{(n)} (\vec{x} , t )  \ .
\end{align}
where $\mathbb{C}_{\mathcal{O}_i, \alpha}^{(n)}$ are the new operators stemming from each operator $\mathcal{O}_i$ upon the Taylor expansion along fluid trajectory. Substituting the form of $\mathcal{O}_i^{(n)}$ from Eq.\eqref{eq:Oin_exp} into Eq.\eqref{eq:tracer_nloc_exp} we get the following bias expansion that is local in time,
\begin{align}\label{eq:nloc_loc_exp}
    \delta^{(n)}_g ( \xvec , t) = \sum_{\mathcal{O}_i} \sum_{\alpha = 1}^{n-i+1} c_{\mathcal{O}_i,\alpha}(t) \,\mathbb{C}^{(n)}_{\mathcal{O}_i,\alpha} ( \xvec , t)\;,
\end{align}
where
\begin{align}\label{eq:nloc_bias_coeff}
    c_{\mathcal{O}_i,\alpha} ( t ) \equiv \int^t dt' c_{\mathcal{O}_i} ( t , t' ) \left( \frac{D(t')}{D(t)} \right)^{\alpha+i-1 } \;,
\end{align}
are the time dependent bias coefficients similar to what is given in the local expansion in Eq.\eqref{eq:tracer_loc_exp}\footnote{In\,\cite{Donath:2023sav}, it has been shown that if the kernels have support over a time scale $1/\omega$ then the non-local in time bias parameters go like $H/\omega$. Therefore, if the bias parameters are order unity, then it implies that the galaxy sample under observation was formed over a timescale of $1/H$.}. From Eq.\eqref{eq:nloc_loc_exp}, it is clear that $\mathbb{C}^{(n)}_{\mathcal{O}_i,\alpha} ( \xvec , t)$ are the local in time operators that affect tracer over density $\delta_g(\xvec,t)$ at location $\xvec$ and time $t$. 

Again we can look for linear relations among different operators appearing in Eq.\eqref{eq:nloc_loc_exp} to get a basis of operators at a given order. Due to Taylor expansion along the fluid trajectory, the structure of operators that appear in  \eqref{eq:nloc_loc_exp} is qualitatively different from those that appear in Eq.\eqref{eq:tracer_loc_exp}. Therefore, in principle, the time local and time non-local expansion can have different basis at each order.

Since the virialisation timescale is small as compared to the timescale of evolution of long modes, we expect that to a certain accuracy, the bias expansion can be taken to be local-in-time\,\cite{Senatore:2014eva}. However, if we go to sufficiently higher order in overdensity, the local in time and the non-local in time bias expansion must differ in their structure. It has been pointed out\,\cite{DAmico:2022ukl,Donath:2023sav}, that up to fourth order in overdensity, non-local expansion can be written in terms of  local expansion i.e the non-local basis and local basis are related by a linear invertible map. 


However, at fifth order, the local and the non-local basis are different and hence, there does not exists a linear map between them\,\cite{Donath:2023sav}. It has been shown that at the fifth order, the time-local basis contains 26 independent operators, while the time non-local basis has 29 independent operators. In fact, at this order the total number of non-local operators that appear in Eq.\eqref{eq:nloc_loc_exp} are of $\mathcal O(100)$ but the operator space has a lot of degeneracies. After removing degeneracies, the number of independent operators turn out to be only 29\,\cite{Donath:2023sav}. Out of the 29 non-local operators, 26 of them can be written as linear combinations of the time-local basis and this map is invertible. This implies that the non-local basis contains three extra operators that cannot be written as a linear combinations of operators in the time-local basis. The three non-local terms are pointed out in\,\cite{Donath:2023sav} and are listed below, 
\begin{align}\label{eq:nloc_terms}
    \Cops{5}{\delta,5},\hspace{0.5cm}\Cops{5}{r^2,4},\hspace{0.5cm}\Cops{5}{p^3,3}\;.
\end{align}
 The presence of these extra non-local terms point towards time non-locality in the physics of structure formation, and it should be possible to detect through data or simulations. In the next sub-section, we discuss the use of a different basis to write non-local expansion, which makes calculation much more efficient and easy. In particular, number of possible structures as well as degeneracies reduces significantly in the new basis as discussed below.

\subsection{$\mathbf{\Pi}$ operators: An efficient way to get the non-local basis}\label{subsec:map_SO(3)}
In this section, we will show that if we use a different system of writing operators, then the task of finding the complete time non-local basis at any give order becomes swift and easy. In Sec.\,\ref{sssec:Pi_basis}, we have described an alternate way of writing operators in the bias expansion\,\cite{Mirbabayi:2014zca}. Taking trace over the contraction of $\Pi_{ij}^{[n]}$'s and their product, we can construct scalars at any give order. We can now perform the same exercise of writing all possible operators relevant at a given order. Then, we can look for degeneracies among them to get the independent set of operators that will form the basis at that order. We have found that up to fifth order, the basis obtained by $\Pi$-system maps directly to the non-local basis through an invertible linear map. At fifth order, we found that there are a total of 52 scalar operators out of which only 29 are independent. The list of 29 independent operators at order five is given as follows,
\begin{equation}\label{eq:Pi_basis}
	\begingroup
	\addtolength{\jot}{0.5em}
	\begin{aligned}
		m = 1 && & \text{Tr}[\Pi^{[1]}] \\
		m = 2 && & \text{Tr}[(\Pi^{[1]})^2],  \; (\text{Tr}[\Pi^{[1]}])^2 \\
		m = 3 && & \text{Tr}[(\Pi^{[1]})^3], \; \text{Tr}[(\Pi^{[1]})^2]\text{Tr}[\Pi^{[1]}],  \; (\text{Tr}[\Pi^{[1]}])^3, \; \text{Tr}[\Pi^{[2]}\Pi^{[1]}]  \\
		m = 4 && & \text{Tr}[(\Pi^{[1]})^4], \; \text{Tr}[(\Pi^{[1]})^3]\text{Tr}[\Pi^{[1]}], \; \left(\text{Tr}[(\Pi^{[1]})^2]\right)^2, \; (\text{Tr}[\Pi^{[1]}])^4, \\
		&& & \text{Tr}[\Pi^{[2]}\Pi^{[1]}\Pi^{[1]}], \; \text{Tr}[\Pi^{[2]}\Pi^{[1]}]\text{Tr}[\Pi^{[1]}], \; \text{Tr}[\Pi^{[2]}\Pi^{[2]}], \; \text{Tr}[\Pi^{[3]}\Pi^{[1]}] \\
        m = 5 && &\text{Tr}[(\Pi^{[1]})^5], \; \text{Tr}[(\Pi^{[1]})^4]\text{Tr}[\Pi^{[1]}], \; \text{Tr}[(\Pi^{[1]})^3](\text{Tr}[\Pi^{[1]}])^2, \; \text{Tr}[(\Pi^{[1]})^3]\text{Tr}[(\Pi^{[1]})^2], \\
		&& & (\text{Tr}[\Pi^{[1]}])^5,\; \text{Tr}[\Pi^{[2]}\Pi^{[1]}\Pi^{[1]}\Pi^{[1]}], \; \text{Tr}[\Pi^{[2]}\Pi^{[1]}\Pi^{[1]}]\text{Tr}[\Pi^{[1]}], \\
		&& & \text{Tr}[\Pi^{[2]}\Pi^{[1]}](\text{Tr}[\Pi^{[1]}])^2,\; \text{Tr}[\Pi^{[2]}\Pi^{[2]}\Pi^{[1]}],\; \text{Tr}[\Pi^{[2]}\Pi^{[2]}]\text{Tr}[\Pi^{[1]}], \; \text{Tr}[\Pi^{[3]}\Pi^{[1]}\Pi^{[1]}], \\
        && & \text{Tr}[\Pi^{[3]}\Pi^{[1]}]\text{Tr}[\Pi^{[1]}],\; \text{Tr}[\Pi^{[3]}\Pi^{[2]}], \; \text{Tr}[\Pi^{[4]}\Pi^{[1]}]\;,
	\end{aligned}
	\endgroup
\end{equation}
where we have arranged the set of operators by the order at which they start and we are denoting it by index m. Also we are using the notation $\text{Tr}[\Pi^{[3]}\Pi^{[1]}]\equiv \Pi^{[3]}_{ij}\Pi^{[1]}_{ij}$. As mentioned earlier, the basis that we get from the $\Pi$ system given in Eq.\eqref{eq:Pi_basis} and from the non-local expansion which is given in Eq.\eqref{eq:nloc_loc_exp} are related by an invertible map up to fifth order. We have found the explicit mapping using \textsc{Mathematica}\footnote{The \textsc{Mathematica} files are attached with the preprint version at arxiv: 2406.17025.}. Here we will just write the three "non-local" operators as pointed in\,\cite{Donath:2023sav} as a linear combination of $\Pi$ operators,
\begin{equation}\label{eq:Pi_so3_op_5}
\begingroup
	\addtolength{\jot}{0.5em}
	\begin{aligned}
		\Cops{5}{\delta,5} &= \frac{2071}{11466}(\text{Tr}[\Pi^{[1]}])^5 - \frac{4022}{9555}\text{Tr}[(\Pi^{[1]})^5] + \frac{591}{3185}\text{Tr}[(\Pi^{[1]})^4]\text{Tr}[\Pi^{[1]}] \\
		\phantom{\mathrel{=}}& \quad + \frac{1303}{5733}\text{Tr}[(\Pi^{[1]})^3](\text{Tr}[\Pi^{[1]}])^2 + \frac{5}{13}\text{Tr}[(\Pi^{[1]})^3]\text{Tr}[(\Pi^{[1]})^2] + \frac{2}{91}\text{Tr}[\Pi^{[2]}\Pi^{[2]}\Pi^{[1]}] \\
	\phantom{\mathrel{=}} & \quad + \frac{3}{130} \text{Tr}[\Pi^{[2]}\Pi^{[2]}]\text{Tr}[\Pi^{[1]}] + \frac{166}{1365}\text{Tr}[\Pi^{[2]}\Pi^{[1]}\Pi^{[1]}\Pi^{[1]}]   \\
	\phantom{\mathrel{=}} & \quad - \frac{8}{1365}\text{Tr}[\Pi^{[2]}\Pi^{[1]}\Pi^{[1]}]\text{Tr}[\Pi^{[1]}] + \frac{2}{13}\text{Tr}[\Pi^{[2]}\Pi^{[1]}](\text{Tr}[\Pi^{[1]}])^2 + \frac{9}{455}\text{Tr}[\Pi^{[3]}\Pi^{[2]}]  \\
	\phantom{\mathrel{=}} & \quad + \frac{38}{1365}\text{Tr}[\Pi^{[3]}\Pi^{[1]}\Pi^{[1]}] + \frac{5}{91}\text{Tr}[\Pi^{[3]}\Pi^{[1]}]\text{Tr}[\Pi^{[1]}] + \frac{2}{39}\text{Tr}[\Pi^{[4]}\Pi^{[1]}] \\
	\Cops{5}{r^2,4} &= 2\text{Tr}[\Pi^{[3]}\Pi^{[2]}] + 4\text{Tr}[\Pi^{[4]}\Pi^{[1]}] \\
	\Cops{5}{p^3,3} &= \frac{204}{175}\text{Tr}[(\Pi^{[1]})^5] + \frac{27}{25}\text{Tr}[\Pi^{[2]}\Pi^{[2]}\Pi^{[1]}] + \frac{432}{175}\text{Tr}[\Pi^{[2]}\Pi^{[1]}\Pi^{[1]}\Pi^{[1]}] \\
	\phantom{\mathrel{=}} & \quad+ \frac{9}{7}\text{Tr}[\Pi^{[3]}\Pi^{[1]}\Pi^{[1]}]\;.
	\end{aligned}
\endgroup
\end{equation}
As mentioned earlier (Eq.\eqref{eq:nloc_terms}), the specific non-local operators cannot be written in terms of the time-local basis but we have shown in Eq.\eqref{eq:Pi_so3_op_5}, they can indeed be written as a linear combination of operators given in the alternative $\Pi$ basis, Eq.\eqref{eq:Pi_basis}. We would like to point out that the $\Pi$ system is more efficient in getting the non-local basis as we just get 52  number of scalar operators which is much less in contrast to  around 130 scalar operators we get from the usual Taylor expansion in Eq.\eqref{eq:nloc_loc_exp}. As a result, at fifth order, it has less number of degenerate operators\footnote{Only 23 degenerate operators are present at fifth order in the $\Pi$ system.} as compared to the non-local expansion which contains $\mathcal{O}(100)$ degenerate operators and hence a lot of redundancies. This reduces the task of finding degeneracy equations which is one of the most challenging aspect of this calculation. Note that Eq.\eqref{eq:Pi_basis} does not contain operators involving velocity gradient. We have checked that the operators constructed from velocity gradient and it's convective derivatives are degenerate with the basis given in Eq.\eqref{eq:Pi_basis} up to fifth order. This was pointed out in\,\cite{Mirbabayi:2014zca}.

In the usual bias expansion, we only consider $SO(3)$ scalars. However, let's (hypothetically) allow for tensors in the bias expansion. Due to the tensorial nature, this will allow for more structure in the operator space. This may lead to time non-locality being shifted to lower orders. One such scenario occurs when we include selection operators in the bias expansion. Selection operators, being a projection of a tensorial object along the line-of-sight, have essentially this tensorial structure which we will discuss in the next section.
 

\section{Non-locality in selection operators
 in the bias expansion}\label{sec:nloc_sel_op}

In this section, we discuss the time non-locality in selection operators which were discussed in Sec.\,\ref{ssec:sel_op}. We have found that including selection operators in the bias expansion shifts the time non-locality to third order instead of fifth order as for the case of $SO(3)$ invariant operators. It has been pointed out in Sec.\ref{ssec:sel_op}, the need for introducing new line of sight dependent operators in the bias expansion especially when writing an EFT of the flux field in Lyman-$\alpha$ forest. We have found that including selection operators in the bias expansion (Eq.\eqref{eq:bias_exp_selec}) the non-locality appears at third order in tracer overdensity. In this section, we will discuss in more detail about this non-locality and show that the time local and time non-local basis differ at third order. 
\subsection{Non-locality shifts to $3^{{\rm rd}}$ order}\label{subsec:nl_3rd_order}
As mentioned earlier, selection operators are needed to take into the line of sight dependent effects that can affect the detection of tracer overdensity. In this section, we discuss the set of independent selection operators appearing at third order. Note that the numbers we quote for independent operators is specific to selection operators only. The full bias expansion, however, will consist of both the $SO(3)$ as well as the selection operators. For the case of Lyman-$\alpha$ forest, all such operators are generated by loops through the exponential map even if they are not present at tree level\,\cite{Ivanov:2023yla}. So we need to include all possible operators with the structure given in Eq.\eqref{eq:sel_op}. Constructing all possible contractions we get the set of local operators. After removing the degenerate operators we find that at third order in tracer overdensity, there are 15 independent local selection operators\footnote{A subset of the operators shown in Eq.\eqref{eq:loc_sel_op_3} are generated when writing the bias expansion in redshift space. However, we have checked that up to third order, the local and the non-local basis in redshift space are equivalent, which has also been noted elsewhere\,\cite{DAmico:2022ukl}.}. The basis of selection operators at third order is given below,
\begin{align}\label{eq:loc_sel_op_3}
	\begingroup
		\addtolength{\jot}{0.5em}
    		\begin{aligned}
    			\left\{  \eta, \; \delta_z, \; \eta^2, \; \delta_z^2, \; \eta\delta, \;  \delta_z\delta, \; (r^2)_\parallel,\;  (p^2)_\parallel, \; \eta^3, \; \eta^2\delta, \; \eta\delta^2,  \; (r^3)_\parallel, \; (r^2)_\parallel\eta,\;  (r^2)_\parallel\delta, \; r^2\eta \right\}
    		\end{aligned}\;,
	\endgroup
\end{align}
where the symbols $\eta$ and $\delta_z$ have been defined in Eq.\eqref{eq:sel_op}.

Based on the same reasoning given in Sec.\,\ref{sec:nloc_bias_exp}, we can argue that the bias expansion in Eq.\eqref{eq:bias_exp_selec} should be generalised to the non-local bias expansion and hence should have integration along the full past trajectory of dark matter fields. Performing the Taylor expansion along the fluid trajectory gives the non-local selection operators with a structure similar to that given in Eq.\eqref{eq:nloc_loc_exp}. After removing the degenerate operators, we find that there are 16 independent non-local selection operators. The list of independent non-local selection operators is
\begin{align}\label{eq:nloc_sel_op_3}
	\begingroup
		\addtolength{\jot}{0.5em}
    		\begin{aligned}
    			 &\left\{ \Cops{3}{\eta,1},\; \Cops{3}{\eta,2},\; \Cops{3}{\eta,3},\; \Cops{3}{\eta^2,1},\; \Cops{3}{\eta^2,2},\; \Cops{3}{\delta_z^2,2},\; \Cops{3}{\eta\delta,1},\; \Cops{3}{\eta\delta,2},\; \Cops{3}{\delta_z\delta,2},\; \Cops{3}{(r^2)_\parallel,1},\; \Cops{3}{(r^2)_\parallel,2},\; \Cops{3}{\eta^3,1},  \right.  \\
    			 \phantom{\mathrel{=}} & \;\; \left.  \Cops{3}{\eta^2\delta,1},\; \Cops{3}{\eta\delta^2,1},\; \Cops{3}{(r^3)_\parallel,1}, \Cops{3}{(r^2)_\parallel\delta,1} \right\}\;.
    		\end{aligned}
	\endgroup
\end{align}
From Eq.\eqref{eq:loc_sel_op_3} and \eqref{eq:nloc_sel_op_3} we can see that the time local basis contains 15 operators while the time non local basis contains 16 operators. This implies that the time non-local basis has one extra operator as compared to the time local basis. Hence, we conclude that including line of sight dependent selection effects enables us to detect non-locality at third order which can be measured in one-loop power spectrum of the $\delta_F$ (Sec.\ref{ssec:sel_op}).

We can also write selection operators in the $\Pi$ system used in Sec.\ref{sssec:Pi_basis}. For that one needs to make two indexed objects from $\Pi_{ij}^{[n]}$ defined in Eq.\eqref{eq:Pi_n} and take projection with the unit vectors along the line of sight such as $(\Pi^{[1]}\Pi^{[2]})_\parallel\equiv\Pi_{ik}^{[1]}\Pi_{kj}^{[2]}\hat{z}^i\hat{z}^j$. At the third order, one finds only 16 independent operators. The list of independent selection operators up to third order in dark matter overdensity is as follows\,\cite{Desjacques:2018pfv},
\begin{align}\label{eq:Pi_sel_op_3}
\begingroup
	\addtolength{\jot}{0.5em}
    \begin{aligned}
        m = 1 && & \Pi_{\parallel}^{[1]} \\
        m = 2 && & \text{Tr}[\Pi^{[1]}] \Pi_{\parallel}^{[1]},\; [ (\Pi^{[1]})^2 ]_{\parallel},\; ( \Pi^{[1]}_{\parallel} )^2,\; \Pi^{[2]}_{\parallel} \\
        m = 3 && & [ (\Pi^{[1]})^3 ]_{\parallel},  \; [(\Pi^{[1]})^2]_{\parallel}\Pi^{[1]}_{\parallel}, \; [(\Pi^{[1]})^2]_{\parallel}\text{Tr}[\Pi^{[1]}], \; \Pi^{[1]}_{\parallel}\text{Tr}[ (\Pi^{[1]})^2 ], \; \Pi^{[1]}_{\parallel}( \text{Tr}[\Pi^{[1]}] )^2,  \\
        && &  (\Pi^{[1]}_{\parallel})^2\text{Tr}[ \Pi^{[1]} ],\; ( \Pi^{[1]}_{\parallel} )^3, \; [ \Pi^{[2]}\Pi^{[1]} ]_{\parallel}, \; \text{Tr}[ \Pi^{[1]} ]\Pi^{[2]}_{\parallel},\; \Pi^{[2]}_{\parallel}\Pi^{[1]}_{\parallel}, \; \Pi^{[3]}_{\parallel} \;,
    \end{aligned}
    \endgroup
\end{align}
where we have arranged the operators by the index ``$m$'' that denotes the order in dark matter overdensity at which a particular operator starts. As was the case for $SO(3)$ operators, we find that here also there exists an invertible map between the non-local basis given in Eq.\eqref{eq:nloc_sel_op_3} and $\Pi$ basis as given in Eq.\eqref{eq:Pi_sel_op_3}. We have found the invertible mapping explicitly using \textsc{Mathematica}. We can see from Eq.\eqref{eq:Pi_sel_op_3}, that it does not contain operators involving velocity gradient. This is due to the fact that the operators constructed from velocity gradient ($\partial_i v_j$) and its convective derivatives are degenerate with the $\Pi$ basis  given in Eq.\eqref{eq:Pi_sel_op_3}. This was pointed out in\,\cite{Mirbabayi:2014zca,Zaldarriaga:2015jrj} and we have explicitly verified this.

The relation between non-local expansion and the $\Pi$ basis holds true at fourth order as well. We will explore this in a bit more detail ahead. As for the case of third order, we can write the \textit{time-local} selection operators up to fourth order by constructing $SO(2)$ scalars. We construct 67 such operators out of which only 40 are independent and form the basis at fourth order. The independent operators are listed below,
\begin{align}\label{eq:loc_sel_4}
 \{&\delta_z,\;\eta,\;\delta_z^2,\;\delta_z\eta,\;\eta^2,\;\delta_z\delta,\;\delta_z\theta,\;\eta\delta,\;\eta\theta,\;(r^2)_\parallel,\;(rp)_\parallel,\;(p^2)_\parallel,\delta_z^3,\;\delta_z^2\eta,\;\delta_z^2\delta,\;\delta_z^2\theta,\;\delta_z\delta^2,\nonumber\\
 &\delta_z\delta\theta,\;\delta_z\eta\delta,\;\eta\delta^2,\;(r^3)_\parallel,\;(r^2p)_\parallel,\;(rp^2)_\parallel,\;(r^2)_\parallel\delta_z,\;(r^2)_\parallel\eta,\;(r^2)_\parallel\delta,\;(r^2)_\parallel\theta,\;(rp)_\parallel\delta_z,\;\nonumber\\
 &(rp)_\parallel\delta,\;r^2\delta_z,\;rp\delta_z,\;\eta^4,\;\eta^3\delta,\;\eta^2\delta^2,\;\eta\delta^3,\;(r^4)_\parallel,\;(r^3)_\parallel\eta,\;(r^2)_\parallel\eta^2,\;(r^2)_\parallel\delta^2,\;((r^2)_\parallel)^2\}\;,  
\end{align}

where we have followed the notation given in Eq.\eqref{eq:SO3_def} and Eq.\eqref{eq:sel_op} for writing operators in a compact form. To get the \textit{time non-local} basis we need to Taylor expand the operators in Eq.\eqref{eq:loc_sel_4} along the fluid trajectory as we did for the case of $SO(3)$ operators given in Eq.\eqref{eq:nloc_loc_exp}. At fourth order, we have found  133 such non-local operators, out of which 88 are degenerate and therefore only 45 operators are independent which we list out here,
\begin{align}\label{eq:nloc_sel_4}
	\begingroup
		\addtolength{\jot}{0.5em}
    		\begin{aligned}
    			&\left\{  \Cops{4}{\eta,1},\; \Cops{4}{\eta^2,1},\; \Cops{4}{\eta\delta,1},\; \Cops{4}{(r^2)_\parallel,1},\; \Cops{4}{\eta^3,1},\; \Cops{4}{\eta^2\delta,1},\; \Cops{4}{\eta\delta^2,1},\; \Cops{4}{(r^3)_\parallel,1},\; \Cops{4}{(r^2)_\parallel\eta,1},\; \Cops{4}{(r^2)_\parallel\delta,1},\; \Cops{4}{r^2\eta,1},\; \Cops{4}{(p^3)_\parallel),2},\right. \\
    			\phantom{\mathrel{=}} & \;\; \left.  \Cops{4}{(prp)_\parallel,2},\; \Cops{4}{(p^2)_\parallel,2},\; \Cops{4}{(p^2)_\parallel,3},\; \Cops{4}{(p^2)_\parallel\delta,2},\; \Cops{4}{(p^2)_\parallel\delta_z,2},\; \Cops{4}{(p^2)_\parallel\eta,2},\; \Cops{4}{(p^2)_\parallel\theta,2},\; \Cops{4}{p^2\delta_z,2},\; \Cops{4}{p^2\eta,2},\; \Cops{4}{(r^3)_\parallel,2},\right. \\
    			\phantom{\mathrel{=}} & \;\; \left.  \Cops{4}{(r^3)_\parallel\delta,1}, \; \Cops{4}{(r^3)_\parallel\eta,1}, \; \Cops{4}{r^3\eta,1},\; \Cops{4}{(rp)_\parallel\delta,2},\; \Cops{4}{(rp)_\parallel\delta_z,2},\; \Cops{4}{(r^2)_\parallel,3},\; \Cops{4}{(r^2)_\parallel\eta^2,1},\; \Cops{4}{(r^2)_\parallel\eta\delta,1},\; \Cops{4}{r^2\eta^2,1},\right. \\ 
    			\phantom{\mathrel{=}} & \;\; \left.  \Cops{4}{r^2\eta\delta,1},\; \Cops{4}{\delta_z,2},\; \Cops{4}{\delta_z,3},\; \Cops{4}{\delta_z,4},\; \Cops{4}{\delta_z^2,2},\; \Cops{4}{\delta_z^2,3},\; \Cops{4}{\delta_z^2\delta,2},\; \Cops{4}{\delta_z^3,2},\right. \\
    			\phantom{\mathrel{=}} & \;\; \left.  \Cops{4}{\delta_z\delta,2},\; \Cops{4}{\delta_z\delta,3},\; \Cops{4}{\delta_z\delta^2,2},\; \Cops{4}{\delta_z\eta,3},\; \Cops{4}{\eta^3\delta,1},\; \Cops{4}{\eta^4,1}  \right\}\;,
    		\end{aligned}
	\endgroup
\end{align}
where we have followed the same notation to write the operators as we used in Eq.\eqref{eq:nloc_loc_exp}.

Now we write the selection operators in the $\Pi$ system. We notice that the 16 operators listed at order three in Eq.\eqref{eq:Pi_sel_op_3} also contribute at fourth order. But now certain new operators appear which start at fourth order in matter overdensity itself. The full list of independent operators at fourth order is given as, 
\begin{align}\label{eq:Pi_sel_4}
\begingroup
	\addtolength{\jot}{0.5em}
    \begin{aligned}
     m = 1 && & \Pi_{\parallel}^{[1]} \\
        m = 2 && & \text{Tr}[\Pi^{[1]}] \Pi_{\parallel}^{[1]},\; [ (\Pi^{[1]})^2 ]_{\parallel},\; ( \Pi^{[1]}_{\parallel} )^2,\; \Pi^{[2]}_{\parallel} \\
        m = 3 && & [ (\Pi^{[1]})^3 ]_{\parallel},  \; [(\Pi^{[1]})^2]_{\parallel}\Pi^{[1]}_{\parallel}, \; [(\Pi^{[1]})^2]_{\parallel}\text{Tr}[\Pi^{[1]}], \; \Pi^{[1]}_{\parallel}\text{Tr}[ (\Pi^{[1]})^2 ], \; \Pi^{[1]}_{\parallel}( \text{Tr}[\Pi^{[1]}] )^2,  \\
        && &  (\Pi^{[1]}_{\parallel})^2\text{Tr}[ \Pi^{[1]} ],\; ( \Pi^{[1]}_{\parallel} )^3, \; [ \Pi^{[2]}\Pi^{[1]} ]_{\parallel}, \; \text{Tr}[ \Pi^{[1]} ]\Pi^{[2]}_{\parallel},\; \Pi^{[2]}_{\parallel}\Pi^{[1]}_{\parallel}, \; \Pi^{[3]}_{\parallel} \\
        m = 4 && & [(\Pi^{[1]})^4]_{\parallel},\; [(\Pi^{[1]})^3]_{\parallel}\Pi^{[1]}_{\parallel},\; \Pi^{[1]}_\parallel\text{Tr}[(\Pi^{[1]})^3],\; [(\Pi^{[1]})^2]_{\parallel}(\text{Tr}[\Pi^{[1]}])^2,\\
        && & [(\Pi^{[1]})^2]_{\parallel}\Pi^{[1]}_{\parallel}\text{Tr}[\Pi^{[1]}],\; [(\Pi^{[1]})^2]_{\parallel}(\Pi^{[1]}_{\parallel})^2,\; (\Pi^{[1]}_{\parallel})^2\text{Tr}[(\Pi^{[1]})^2], \; (\Pi^{[1]}_{\parallel})^4,\\
        && & (\Pi^{[1]}_{\parallel})^3\text{Tr}[\Pi^{[1]}],\; (\Pi^{[1]}_{\parallel})^2(\text{Tr}[\Pi^{[1]}])^2,\; \Pi^{[1]}_{\parallel}(\text{Tr}[\Pi^{[1]}])^3,\;[(\Pi^{[1]})^3]_\parallel\text{Tr}[\Pi^{[1]}],\\
        && &([(\Pi^{[1]})^2]_{\parallel})^2,\; [\Pi^{[2]}\Pi^{[1]}\Pi^{[1]}]_{\parallel},\; [\Pi^{[1]}\Pi^{[2]}\Pi^{[1]}]_{\parallel},\;[\Pi^{[2]}\Pi^{[1]}]_{\parallel}\Pi^{[1]}_{\parallel},\; [\Pi^{[2]}\Pi^{[1]}]_{\parallel}\text{Tr}[\Pi^{[1]}],\\
        && & \Pi^{[1]}_{\parallel}\text{Tr}[\Pi^{[2]}\Pi^{[1]}],\; \Pi^{[2]}_{\parallel}[(\Pi^{[1]})^2]_{\parallel},\; \Pi^{[2]}_{\parallel}\text{Tr}[(\Pi^{[1]})^2],\; \Pi^{[2]}_{\parallel}(\Pi^{[1]}_{\parallel})^2,\; \Pi^{[2]}_{\parallel}\Pi^{[1]}_{\parallel}\text{Tr}[\Pi^{[1]}],\\
        && & \Pi^{[2]}_{\parallel}(\text{Tr}[\Pi^{[1]}])^2,\; [\Pi^{[2]}\Pi^{[2]}]_{\parallel},\; (\Pi^{[2]}_{\parallel})^2,\; [\Pi^{[3]}\Pi^{[1]}]_{\parallel}, \; \Pi^{[3]}_{\parallel}\text{Tr}[\Pi^{[1]}],\; \Pi^{[3]}_{\parallel}\Pi^{[1]}_{\parallel}, \; \Pi^{[4]}_{\parallel}\;.
    \end{aligned}
    \endgroup
\end{align}
So Eq.\eqref{eq:Pi_sel_4} together gives the total number of selection operators at fourth order. From Eq.\eqref{eq:Pi_sel_4} we can see that there are 29 new operators that appear at fourth order coming from $m=4$ sector, which makes the total number of selection operators at order four to be 45.

The $\Pi$ operators given in Eq.\eqref{eq:Pi_sel_4} are related to the time non-local basis with an invertible map. We have found the explicit mapping using \textsc{Mathematica}.\footnote{Note that Eq.\eqref{eq:Pi_sel_4} does not contain operators with velocity gradient ($\partial_iv_j$) and it's convective derivatives. This is because at fourth order also, all such operators are degenerate with the operators given in Eq.\eqref{eq:Pi_sel_4}. At fourth order, we can write 53 operators involving velocity gradient and their convective derivatives. We have explicitly checked that each of them is degenerate with the $\Pi$ operators given in Eq.\eqref{eq:Pi_sel_4}.}

\subsection{Connection with observation}
We have found through Eq.\eqref{eq:loc_sel_op_3} and \eqref{eq:nloc_sel_op_3}, that the time non-local basis at third order contains one more operator than the time local basis. This implies that there exists one operator in the non-local basis that cannot be written in terms of the local basis. We have also shown that the time non-local basis is mathematically equivalent to the $\Pi$ basis given in Eq.\eqref{eq:Pi_sel_op_3}. Among the operators given in Eq.\eqref{eq:Pi_sel_op_3}, all operators except $\Pi_\parallel^{[2]}$ and $\Pi_\parallel^{[3]}$ are writable in terms of the time local basis given in Eq.\eqref{eq:loc_sel_op_3}. Interestingly, the linear combination $7(\Pi^{[2]}_\parallel)^{(3)}-4(\Pi^{[3]}_\parallel)^{(3)}$ is writable in terms of the local basis given in Eq.\eqref{eq:loc_sel_op_3}, revealing that this combination is time local. However, any other linear combination such as,
\begin{align}\label{eq:non_loc_op_3}
\mathcal{O}_{\text{non-loc}}^{(3)}=7(\Pi^{[2]}_\parallel)^{(3)}+4(\Pi^{[3]}_\parallel)^{(3)}\;,
\end{align} 
is not writable in terms of the time local basis and hence, is intrinsically time non-local. Following\,\cite{Desjacques:2018pfv}, we show that the time non-local operator given in Eq.\eqref{eq:non_loc_op_3} gives non trivial contribution to the 1-loop power spectrum.

Let us give a brief overview of the basic idea given in\,\cite{Desjacques:2018pfv}. The power spectrum upto one loop has the following structure,
\begin{align}
\langle\delta_\text{\small{F}}(\boldsymbol{k}_1,\eta)\delta_\text{\small{F}}(\boldsymbol{k}_2,\eta)\rangle=&D^2\langle\delta_\text{\small{F}}^{(1)}(\boldsymbol{k}_1)\delta_\text{\small{F}}^{(1)}(\boldsymbol{k}_2)\rangle\nonumber\\
&+D^4(\langle\delta_\text{\small{F}}^{(2)}(\boldsymbol{k}_1)\delta_\text{\small{F}}^{(2)}(\boldsymbol{k}_2)\rangle+2\langle\delta_\text{\small{F}}^{(1)}(\boldsymbol{k}_1)\delta_\text{\small{F}}^{(3)}(\boldsymbol{k}_2)\rangle)\;,
\end{align}
where $D$ is the linear growth factor given in Eq.\eqref{eq:ansatz_series}. We will only be analysing the contribution coming from $\langle\delta_\text{\small{F}}^{(1)}(\boldsymbol{k}_1)\delta_\text{\small{F}}^{(3)}(\boldsymbol{k}_2)\rangle$, as time non-locality manifests at third order. The general structure of this contribution has the following form,
\begin{align}\label{eq:cont_O}
\langle\delta^{(1)}(\boldsymbol{k}_1)\mathcal{O}^{(3)}(\boldsymbol{k}_2)\rangle\;,
\end{align}
where $\mathcal{O}^{(3)}$ is one of the operators from the list given in Eq.\eqref{eq:Pi_sel_op_3}. Any cubic operator can be expressed as,
\begin{align}\label{eq:O_form}
    \mathcal{O}^{(3)}(k)=\int_{\boldsymbol{q}_1\boldsymbol{q}_2\boldsymbol{q}_3}\delta^{\text{D}}(\boldsymbol{k}-\boldsymbol{q}_{123})F_{\mathcal{O}}(\boldsymbol{q}_1,\boldsymbol{q}_2,\boldsymbol{q}_3)\delta^{(1)}_{\boldsymbol{q}_1}\delta^{(1)}_{\boldsymbol{q}_2}\delta^{(1)}_{\boldsymbol{q}_3}\;.
\end{align}
Using Eq.\eqref{eq:O_form} in Eq.\eqref{eq:cont_O} we get the most general structure of the 1-3 contribution,
\begin{align}\label{eq:gen_str_13_cont}
    \langle\delta^{(1)}(\boldsymbol{k}_1)\mathcal{O}^{(3)}(\boldsymbol{k}_2)\rangle=\int_{\boldsymbol{q}_1\boldsymbol{q}_2\boldsymbol{q}_3}\delta^{\text{D}}(\boldsymbol{k}_2-\boldsymbol{q}_{123})F_{\mathcal{O}}(\boldsymbol{q}_1,\boldsymbol{q}_2,\boldsymbol{q}_3)\langle\delta^{(1)}_{\boldsymbol{k}_1}\delta^{(1)}_{\boldsymbol{q}_1}\delta^{(1)}_{\boldsymbol{q}_2}\delta^{(1)}_{\boldsymbol{q}_3}\rangle
\end{align}
Let us now look at the form of the possible kernels appearing in Eq.\eqref{eq:gen_str_13_cont}. One class of operators, that contains products of three $\Pi^{[1]}_{ij}$, have kernel of the following form\,\cite{Desjacques:2018pfv}, 
\begin{align}\label{eq:F1}
    F_\mathcal{O}(\boldsymbol{q}_1,\boldsymbol{q}_2,\boldsymbol{q}_3)=\mathcal{P}^{\mathcal{O}}_{ijklmn}\frac{q_1^iq_1^j}{q_1^2}\frac{q_2^kq_2^l}{q_2^2}\frac{q_3^mq_3^n}{q_3^2}
\end{align}
where $\mathcal{P}^{\mathcal{O}}$ is a projection operator made out of products of Kronecker delta and even powers of $\hat{z}^i$. Operators whose kernels take the form of Eq.\eqref{eq:F1} have been shown to be fully absorbed by counterterms, and therefore do not contribute independently to the 1-loop power spectrum\,\cite{Desjacques:2018pfv}.
Another class of kernels that appear at third order is given as,
\begin{align}\label{eq:F2}
     F_\mathcal{O}(\boldsymbol{q}_1,\boldsymbol{q}_2,\boldsymbol{q}_3)=\mathcal{P}'^{\mathcal{O}}_{ijkl}\frac{q_1^iq_1^j}{q_1^2}\delta^{[1]}\frac{q_{23}^kq_{23}^l}{q_{23}^2}(\mathcal{O}^{[2]})\;,
\end{align}
where the projection operator $\mathcal{P}'^{\mathcal{O}}_{ijkl}$ is again made out of products of Kronecker delta and even powers of $\hat{z}^i$ and in general, is different from the one given in Eq.\eqref{eq:F1}. However, if it contains $\delta_{kl}$ then Eq.\eqref{eq:F2} takes the same form as Eq.\eqref{eq:F1} and hence do not contribute to the 1-loop power spectrum. For more details see Appendix.\,C and D of\,\cite{Desjacques:2018pfv}.

The explicit form for both the operators given in Eq.\eqref{eq:non_loc_op_3} is as follows,
\begin{align}\label{eq:pi2_pi3_para}
(\Pi^{[2]}_\parallel)^{(3)}&=\left(\frac{2\partial_i\partial_j}{\partial^2}\delta^{(3)}-\frac{\partial_k}{\partial^2}\theta^{(1)}\frac{\partial_k\partial_i\partial_j}{\partial^2}\delta^{(2)}-\frac{\partial_k}{\partial^2}\theta^{(2)}\frac{\partial_k\partial_i\partial_j}{\partial^2}\delta^{(1)}\right)\hat{z}^i\hat{z}^j\nonumber\\
(\Pi^{[3]}_\parallel)^{(3)}&=\left(\frac{\partial_i\partial_j}{\partial^2}\delta^{(3)}-2\frac{\partial_k}{\partial^2}\theta^{(1)}\frac{\partial_k\partial_i\partial_j}{\partial^2}\delta^{(2)}-\frac{\partial_k}{\partial^2}\theta^{(2)}\frac{\partial_k\partial_i\partial_j}{\partial^2}\delta^{(1)}\right.\nonumber\\
&\left.+\frac{\partial_k}{\partial^2}\theta^{(1)}\frac{\partial_l}{\partial^2}\theta^{(1)}\frac{\partial_k\partial_l\partial_i\partial_j}{\partial^2}\delta^{(1)}+\frac{\partial_l}{\partial^2}\theta^{(1)}\frac{\partial_l\partial_k}{\partial^2}\theta^{(1)}\frac{\partial_k\partial_i\partial_j}{\partial^2}\delta^{(1)}\right)\hat{z}^i\hat{z}^j\;.
\end{align}
The last line of Eq.\eqref{eq:pi2_pi3_para} is of the form given in Eq.\eqref{eq:F1}, which implies that it can be absorbed by counterterms and does not contribute to the 1-loop power spectrum. However, all other terms in Eq.\eqref{eq:pi2_pi3_para} give non-trivial contributions to the 1-loop power spectrum individually as well as in the linear combination given in Eq.\eqref{eq:non_loc_op_3}. Hence, we conclude that the time non-local operator given in Eq.\eqref{eq:non_loc_op_3} is relevant at one-loop power spectrum. This suggests that time non-local effects can be measured in the power spectrum of $\delta_F$ at 1-loop through a contribution that goes like $\langle\delta_F^{(3)}(\boldsymbol{k}_1)\delta_F^{(1)}(\boldsymbol{k}_2)\rangle$. At tree level, time non-locality will affect the four-point function $\langle\delta_F^{(3)}(\boldsymbol{k}_1)\delta_F^{(1)}(\boldsymbol{k}_2)\delta_F^{(1)}(\boldsymbol{k}_3)\delta_F^{(1)}(\boldsymbol{k}_4)\rangle$ as well.




\section{Discussion and Conclusion}\label{sec:disc_conc}
Galaxy formation time is of order $(1/10)^{\text{th}}$ the evolution timescales of the Universe as a whole and hence is inherently a time non-local process. Indeed, in the context of the EFTofLSS, \cite{Donath:2023sav} has recently shown that time non-locality arises explicitly at the fifth order in bias expansion. However, a natural question is that if galaxy formation is intrinsically time-non-local in nature, why does this aspect show up only at the fifth order in the expansion? Our analysis reveals that delving beyond $SO(3)$ scalar observables, that were considered in \cite{Donath:2023sav}, can unveil non-locality even at lower orders. For instance, in the context of the effective stress tensor within the framework of EFTofLSS --- $\tau_{ij}$ --- which transforms as a second-rank tensor under $SO(3)$ transformations, temporal non-locality manifests at the third order\footnote{This non-locality in dark matter is anticipated due to the lack of separation between the time scales of long and short modes.}. This implies that one necessarily has to use time non-local counterterms in the effective stress tensor beyond second order. However, in any calculation involving stress tensor only up to second order, one can simply use the local counterterms which are simpler to work with. Another instance where time non-locality is indispensable arises when incorporating selection operators in the bias expansion. These operators, acting as projections of tensorial objects along the line-of-sight, inherently possess the structures necessary to exhibit  temporal non-locality at lower orders. Notably, in the case of Lyman-$\alpha$ flux fluctuations, where the symmetry breaks down to SO(2) along the line of sight, temporal non-locality emerges again at the third order in the bias expansion\footnote{Let us note that in \cite{Donath:2023sav} and in the current paper, we have only used SPT kernels which corresponds to non-renormalised matter overdensity and velocity divergence as given in\,\eqref{eq:kernels}.}. This suggests that temporal non-locality might be observable in the 1-loop power spectrum for phenomena necessitating the inclusion of selection operators. This is in contrast to the $SO(3)$ bias expansion, where non-locality is only detectable at the two-loop power spectrum level for observables. Therefore, we see that time non-locality manifests itself at different orders for different observables. Our analysis has pointed out this distinction.

In addition to exploring these physics concepts, our paper delved into several intriguing technical aspects. Typically, non-locality emerges at significantly higher orders in perturbative expansions. For instance, in the case of the $SO(3)$ bias expansion, it arises at the fifth order. Analyzing non-locality at these higher orders presents considerable technical challenges. Therefore, gaining insight into more efficient methods for investigating non-locality is crucial.
In this study, we discovered that employing the $\Pi$ basis\,\cite{Mirbabayi:2014zca} offers a more efficient approach to handling non-locality. The novel aspect of our study is to identify momentum structures in $\Pi$ (time non-local) basis such as those given in Eq.\eqref{eq:non_loc_op_3}, which contains inherently non-local-in-time physics. The method to generate different bases (time local, non-local and $\Pi$), however, is not new and has been the object of many studies\,\cite{Senatore:2014eva,Mirbabayi:2014zca,DAmico:2022ukl,Donath:2023sav,Abolhasani:2015mra,Vlah:2019byq}. Specifically, concerning biased tracers, we demonstrated that the correspondence between the $\Pi$ basis and the time non-local basis extends up to the fifth order for $SO(3)$ invariant operators. It's noteworthy that the $\Pi$ system proves more effective in establishing the non-local basis, requiring only 52 scalar operators at fifth order to begin with, a significant reduction compared to the approximately 130 scalar operators in the conventional approach.
We have also established the efficacy of the $\Pi$ basis when dealing with selection operators. 

In deriving these results, we have employed the usual method of writing selection operators in the bias expansion as given in Eq.\eqref{eq:sel_op}. In this method we write all possible $SO(2)$ invariant operators at a given order. This is termed as the bottom-up approach to writing operators in the bias expansion. However, for the case of Lyman-$\alpha$ forest, we can generate the same set of operators in the bias expansion of flux fluctuation ($\delta_F$) through renormalisation of the flux field. Due to the non-linear exponential map between the optical depth ($\tau$) and the total flux ($F$), given as $F=e^{-\tau}$, $\delta_F$ contains operators with products of $\delta_{\tau}$ evaluated at the same point in space. This requires renormalisation of these composite operators. Taking the optical depth fluctuation in redshift space and renormalising the $\delta_F$ field, in principle, should generate all possible operators that appear in the bias expansion in Eq.\eqref{eq:sel_op}. For more details see Appendix.\,\ref{ap:renorm}.

Below we list out some of the future directions which we think would be very interesting to pursue.

\begin{itemize}
\item
In\,\cite{Assassi:2014fva}, it has been pointed out that renormalisation generates all operators that are allowed by the symmetries. It would be intriguing to check, whether renormalisation generates the time local or time non-local operators for the case of both $SO(3)$ and $SO(2)$ invariant operators.

\item In order to derive the non-local basis for biased tracers, no assumption is made regarding whether tracers should follow continuity equation or not\,\cite{Donath:2023sav}. $\Pi$ system, on the other hand, is derived by assuming that the tracers satisfy a continuity equation. However, we have shown that the non-local basis is related to the $\Pi$ basis by an invertible map up to fifth for $SO(3)$ invariant operators and up to fourth order for $SO(2)$ invariant operators. It would be interesting to have a physical understanding of this mathematical equivalence.

\item One may wonder whether the equivalence of the $\Pi$ and the time non-local basis extends beyond fifth order. Doing explicit computation of the basis and comparing will be very challenging as we know from fifth order. It would be nice to have a formal proof of this equivalence between the $\Pi$ and the time non-local basis at all orders. There is already a hint to why this should happen given the fact that $\Pi$ operators given in Eq.\eqref{eq:Pi_n} are just Lagrangian space operators derived by Taylor expanding the Eulerian operators about the initial time slice\,\cite{Mirbabayi:2014zca}. This Taylor expansion is similar to the Taylor expansion we do to obtain the time non-local operators as given in Eq.\eqref{eq:Oin_exp_dm}.

\item Presence of time non-locality in the effective stress tensor as given in Table.\,\ref{tab:dm_nloc}, implies that the structure of kernels for dark matter overdensity as given in Eq.\eqref{eq:kernels} should also change after renormalisation as these new non-local terms will be added as counterterms to the dark matter overdensity field\,\cite{DAmico:2022ukl}. As a result, if we write the operators in bias expansion with the renormalised kernels of EFT rather than using SPT kernels as given in Eq.\eqref{eq:kernels}, it would be interesting to see how the analysis of Secs.\,\ref{sec:nloc_bias_exp} and \ref{sec:nloc_sel_op} will be modified. We leave this investigation for future exploration. Also in this study, we have not considered higher derivative operators in the bias expansion. It would be interesting to see how the results are modified when higher derivative operators are also considered for both $SO(3)$ and selection operators. Given the fact that time non-locality appears at fourth order in $\partial_i\partial_j\tau_{ij}$, which is a two derivative operator, one would expect the same for higher derivative operators. Exploring this aspect of bias expansion is left for future work. 
\item Recently, bootstrap ideas have been applied to write the kernels for the biased tracers\,\cite{DAmico:2021rdb}. Just using symmetry properties of the equations of motion, it has been shown that the kernels for biased tracers can be bootstrapped up to third order. It would be interesting to investigate time non-locality from a bootstrap perspective. Preliminary investigation along these lines suggests one interesting technical point, which is the presence of an even more efficient basis than the $\Pi$ basis that we have employed in this paper.

\item One can also do a full $N$-body simulation and measure the contribution of time non-local operators found in this study and the studies preceding this\,\cite{Donath:2023sav}. An exhaustive $N$-body simulation with generic initial conditions maybe computationally challenging. However, with simpler initial conditions, see for e.g.\,\cite{Karandikar:2023ozp}, such computations maybe tractable. Measuring the contributions due to time non-local operators in simulations will establish time non-locality as a generic feature of LSS.

\end{itemize}


\acknowledgments
AA acknowledges the support from a
Senior Research Fellowship, granted by the Human Resource Development Group, Council
of Scientific and Industrial Research, Government of India. AB acknowledges support from Science and Engineering Research Board (SERB) India via the Startup Research Grant SRG/2023/000378. SJ would like to thank participant of Cosmology Journal club of iiser pune and DTP, TIFR seminar series for their probing questions and which helped clarify many important physics issues. SJ also would like to thank S. Minwalla, B. Dasgupta for their insightful discussions. We would also like to thank B. Dasgupta for useful comments on an earlier version of the draft. The authors would like to thank M. Lewandowski for pointing out an error and  useful comments on an earlier version of the manuscript. The authors would also like to acknowledge their debt to the people of India for their steady support of research in basic sciences.

\appendix
\section{Basis of vector operators ($\partial_j\tau_{ij}$) up to third order}\label{ap:vec_op}
In Sec.\,\ref{sec:nloc_dm}, we have mentioned about time non-locality appearing in tensor ($\tau_{ij}$) and vector ($\partial_j\tau_{ij}$) operators. We have given the local, non-local and $\Pi$ basis appearing in the tensor sector upto third order in Table.\,\ref{tab:dm_nloc}. Here we provide local, non-local and $\Pi$ basis for the vector operators.

At linear order in matter overdensity there is only one independent operator in the local, non-local  as well as $\Pi$ system. We list here the independent set of operators,
\begin{align}\label{eq:vec_op_1}
    \text{Local basis:} &\hspace{0.5cm}\partial_jr_{ij},\hspace{0.5cm} \nonumber\\
    \text{Non-local basis:}&\hspace{0.5cm}\Cops{1}{\partial_jr_{ij},1},\hspace{0.5cm}\nonumber\\
    \Pi\text{-basis:}&\hspace{0.3cm}\partial_j\Pi_{ij}^{[1]}
\end{align}
where we have used the same notation as is used in writing the operators in Table.\,\ref{tab:dm_nloc}.

\textbf{Second Order}

At second order also, the local, non-local and $\Pi$ basis each have four operators which are listed below,
\begin{align}\label{eq:vec_op_2}
    \text{Local basis:} &\hspace{0.3cm}\{\partial_j(r_{ij}\delta), \partial_i(\delta, \delta^2,r^2)\}\nonumber\\
    \text{Non-local basis:}&\hspace{0.3cm}\{\Cops{2}{\partial_jp_{ij},1},\Cops{2}{\partial_i\delta,1},\Cops{2}{\partial_i\delta,2},\Cops{2}{\partial_i\theta,2}\}\nonumber\\
    \Pi\text{-basis:}&\hspace{0.3cm}\{\partial_j(\Pi_{ij}^{[1]}\text{Tr}[\Pi^{[1]}]),\partial_i(\text{Tr}[\Pi^{[1]}],\text{Tr}[\Pi^{[1]}]^2,\text{Tr}[\Pi^{[1]}\Pi^{[1]}])\}\;.
\end{align}

\textbf{Third order}

At third order, the local basis consists of 12 operators while the non-local basis contains 13 independent operators. Hence, we find that the local and non-local basis differ from each other. As expected, we find 13 independent operators in the $\Pi$ basis. We list out the set of operators that appear in the three basis below,
\begin{align}\label{eq:vec_op_3}
    \text{Local basis:} \hspace{0.3cm}&\{\partial_j(p_{ij}\delta,p_{ij}\theta,r_{ij}\delta,p_{ij}\delta^2,(r^3)_{ij}),\partial_i(\delta,\theta,\delta^2,\delta\theta,r^2,rp,\delta^3)\}\nonumber\\
    \text{Non-local basis:}\hspace{0.3cm}&\{\Cops{3}{\partial_jr_{ij},1},\Cops{3}{\partial_jr_{ij},2},\Cops{3}{\partial_jp_{ij},2},\Cops{3}{\partial_j(p_{ij}\delta),1},\Cops{3}{\partial_j(p_{ij}\delta),2},\Cops{3}{\partial_j(r_{ij}\delta),2},\nonumber\\
    &\Cops{3}{\partial_i\delta,1},\Cops{3}{\partial_i\delta,2},\Cops{3}{\partial_i\delta,3},\Cops{3}{\partial_i\theta,2},\Cops{3}{\partial_i\theta,3},\Cops{3}{\partial_i\delta^2,2},\Cops{3}{\partial_ir^2,2}\}\nonumber\\
    \Pi\text{-basis:}\hspace{0.3cm}&\{\partial_j(\text{Tr}[\Pi^{[1]}]\Pi_{ij}^{[1]},\Pi_{ij}^{[2]},\Pi_{ij}^{[1]}\text{Tr}[\Pi^{[1]}]^2,[\Pi^{[1]}\Pi^{[1]}\Pi^{[1]}]_{ij},[\Pi^{[1]}\Pi^{[1]}]_{ij}\text{Tr}[\Pi^{[1]}],\nonumber\\
    &\text{Tr}[\Pi^{[1]}]\Pi_{ij}^{[2]}),\nonumber\\
    &\partial_i(\text{Tr}[\Pi^{[1]}],\text{Tr}[\Pi^{[1]}]^2,\text{Tr}[\Pi^{[1]}\Pi^{[1]}],\text{Tr}[\Pi^{[2]}],\text{Tr}[\Pi^{[1]}]^3,\text{Tr}[\Pi^{[1]}\Pi^{[1]}\Pi^{[1]}],\nonumber\\
    &\text{Tr}[\Pi^{[1]}\Pi^{[1]}]\text{Tr}[\Pi^{[1]}])\}
\end{align}

From Eq.\eqref{eq:vec_op_3}, we can see that the non-local basis contains 13 while the local basis contains only 12 operators. Hence, there cannot exist an invertible map between them. Also we see that the $\Pi$ basis contains the same number of operators as the non-local basis. We have explicitly checked that there is an invertible map between the non-local and $\Pi$ basis. 

The extra term in the non-local basis coming in the $\partial_j\tau_{ij}$ will contribute in the the equation of motion given in Eq.\eqref{eq:fluid_eq_long}\,\cite{DAmico:2022ukl}. Hence, the kernels for $v^i$ will change as compared to the case when the local vector operator is considered.
\section{Basis of scalar operators ($\partial_i\partial_j\tau_{ij}$) up to fourth order}\label{ap:scal_ops}
In Sec.\,\ref{sec:intro}, we have stated that for scalar operators ($\partial_i\partial_j\tau_{ij}$), time non-locality manifests at fourth order. To demonstrate the same, we explicitly state the set of local, non-local and $\Pi$ basis for scalar operators upto fourth order. We will show that upto third order all systems form equivalent bases which are related by linear invertible maps. However, at fourth order, time non-local basis differs from time local basis and hence, no map exists between time local and non-local basis. Nevertheless, $\Pi$ basis is related to time non-local basis at each order. Note that the operators appearing in $\partial_i\partial_j\tau_{ij}$ are also relevant for higher derivative operators in galaxy bias.

Below, we list the independent set of operators at each order sequentially. At linear order in matter overdensity, there is only one independent operator in the local, non-local, as well as $\Pi$ system,
\begin{align}\label{eq:scala_op_1}
    \text{Local basis:} &\hspace{0.5cm}\partial_i\partial_jr_{ij},\hspace{0.5cm} \nonumber\\
    \text{Non-local basis:}&\hspace{0.5cm}\Cops{1}{\partial_i\partial_jr_{ij},1},\hspace{0.5cm}\nonumber\\
    \Pi\text{-basis:}&\hspace{0.3cm}\partial_i\partial_j\Pi_{ij}^{[1]}\;,
\end{align}
where we have used the same convention for writing time non-local operators as used in Sec.\,\ref{sec:nloc_bias_exp}.

\textbf{Second Order}

At second order, we get 4 independent scalar operators in the time local, non-local, and $\Pi$ basis. The set of independent operators are as follows,
\begin{align}\label{eq:scala_op_2}
    \text{Local basis:} &\hspace{0.3cm}\{\partial_i\partial_j(r_{ij}\delta), \partial^2(\delta, \delta^2,r^2)\}\nonumber\\
    \text{Non-local basis:}&\hspace{0.3cm}\{\Cops{2}{\partial_i\partial_jp_{ij},1},\Cops{2}{\partial^2\delta,1},\Cops{2}{\partial^2\delta,2},\Cops{2}{\partial^2\theta,2}\}\nonumber\\
    \Pi\text{-basis:}&\hspace{0.3cm}\{\partial_i\partial_j(\Pi_{ij}^{[1]}\text{Tr}[\Pi^{[1]}]),\partial^2(\text{Tr}[\Pi^{[1]}],\text{Tr}[\Pi^{[1]}]^2,\text{Tr}[\Pi^{[1]}\Pi^{[1]}])\}\;.
\end{align}
We have explicitly checked that all three bases given in Eq.\eqref{eq:scala_op_2} are related by linear maps to each other.

\textbf{Third Order}

At third order, as in the case of vector operators, we have 12 independent scalar operators in the local basis. In case of vector operators, time non-local and $\Pi$ basis contains 13 operators as given in Eq.\eqref{eq:vec_op_3}. However, in contrast to the case of vector operators, scalar operators in time non-local and $\Pi$ basis contain 12 independent operators which are given as,
\begin{align}\label{eq:scala_op_3}
    \text{Local basis:} \hspace{0.3cm}&\{\partial_i\partial_j(p_{ij}\delta,p_{ij}\theta,r_{ij}\delta,p_{ij}\delta^2,(r^3)_{ij}),\partial^2(\delta,\theta,\delta^2,\delta\theta,r^2,rp,\delta^3)\}\nonumber\\
    \text{Non-local basis:}\hspace{0.3cm}&\{\Cops{3}{\partial_i\partial_jr_{ij},1},\Cops{3}{\partial_i\partial_jr_{ij},2},\Cops{3}{\partial_i\partial_jp_{ij},2},\Cops{3}{\partial_i\partial_j(p_{ij}\delta),1},\Cops{3}{\partial_i\partial_j(p_{ij}\delta),2},\nonumber\\
    &\Cops{3}{\partial^2\delta,1},\Cops{3}{\partial^2\delta,2},\Cops{3}{\partial^2\delta,3},\Cops{3}{\partial^2\theta,2},\Cops{3}{\partial^2\theta,3},\Cops{3}{\partial^2\delta^2,2},\Cops{3}{\partial^2r^2,2}\}\nonumber\\
    \Pi\text{-basis:}\hspace{0.3cm}&\{\partial_i\partial_j(\text{Tr}[\Pi^{[1]}]\Pi_{ij}^{[1]},\Pi_{ij}^{[1]}\text{Tr}[\Pi^{[1]}]^2,[\Pi^{[1]}\Pi^{[1]}\Pi^{[1]}]_{ij},[\Pi^{[1]}\Pi^{[1]}]_{ij}\text{Tr}[\Pi^{[1]}],\text{Tr}[\Pi^{[1]}]\Pi_{ij}^{[2]}),\nonumber\\
    &\partial^2(\text{Tr}[\Pi^{[1]}],\text{Tr}[\Pi^{[1]}]^2,\text{Tr}[\Pi^{[1]}\Pi^{[1]}],\text{Tr}[\Pi^{[2]}],\text{Tr}[\Pi^{[1]}]^3,\text{Tr}[\Pi^{[1]}\Pi^{[1]}\Pi^{[1]}],\nonumber\\
    &\text{Tr}[\Pi^{[1]}\Pi^{[1]}]\text{Tr}[\Pi^{[1]}])\}\;.
\end{align}
We have checked that the different bases given in Eq.\eqref{eq:scala_op_3} are related by invertible linear maps. Hence, there are no non-trivial consequences of time non-locality at third order for scalar operators. However, as we will see below, this will change at fourth order.

\textbf{Fourth Order}

At fourth order, there are 30 independent scalar operators in the time local basis, whereas, the time non-local and $\Pi$ basis contains 31 independent operators. This implies that the time non-locality has non-trivial effects at fourth order. Below, we list the set of independent operators at time local, non-local and $\Pi$ basis,
\begin{align}\label{eq:scala_op_4}
    \text{Local basis:} \hspace{0.3cm}&\{\partial_i\partial_j(r_{ij}\delta,r_{ij}\theta,p_{ij}\delta,p_{ij}\theta,r_{ij}\delta^2,r_{ij}\delta\theta,p_{ij}\delta^2,(r^3)_{ij},(r^2p)_{ij},\nonumber\\
    &(rp^2)_{ij},(r)^2_{ij}\delta,(r)^2_{ij}\theta,(rp)_{ij}\delta,p_{ij}\delta^3,(r)^4_{ij})\nonumber\\
&\partial^2(\delta,\theta,\delta^2,\delta\theta,\theta^2,r^2,rp,p^2,\delta^3,\delta^2\theta,r^3,r^2p,r^2\delta,rp\delta,\delta^4)\}\nonumber\\
    \text{Non-local basis:}\hspace{0.3cm}&\{\Cops{4}{\partial_i\partial_jp_{ij},1},\Cops{4}{\partial_i\partial_jp_{ij}\delta,1},\Cops{4}{\partial_i\partial_j(r)^2_{ij},1},\Cops{4}{\partial_i\partial_j(p_{ij}\delta^2),1},\Cops{4}{\partial_i\partial_j((r^3)_{ij}),1},\Cops{4}{\partial_i\partial_j(r^2_{ij}\delta),1},\nonumber\\
    &\Cops{4}{\partial_i\partial_j(p_{ij}\delta^3),1},\Cops{4}{\partial_i\partial_j(r^4_{ij}),1},\Cops{4}{\partial_i\partial_j(r^3_{ij}\delta),1},\Cops{4}{\partial_i\partial_j(r^3p_{ij}),1},\Cops{4}{\partial_i\partial_j(r^2_{ij}\delta^2),1},\Cops{4}{\partial_i\partial_j(p^3_{ij}),2},\nonumber\\
    &\Cops{4}{\partial_i\partial_j((prp)_{ij}),2},\Cops{4}{\partial_i\partial_j(p^2_{ij}),2},\Cops{4}{\partial_i\partial_j(p^2_{ij}),3},\Cops{4}{\partial_i\partial_j(p^2r_{ij}),2},\nonumber\\
    &\Cops{4}{\partial^2\delta,1},\Cops{4}{\partial^2\delta,2},\Cops{4}{\partial^2\delta,3},\Cops{4}{\partial^2\delta,4},\Cops{4}{\partial^2\theta,2},\Cops{4}{\partial^2\theta,3},\Cops{4}{\partial^2\theta,4},\Cops{4}{\partial^2\delta^2,2},\Cops{4}{\partial^2\delta^2,3},\nonumber\\
    &\Cops{4}{\partial^2\theta^2,3},\Cops{4}{\partial^2r^2,2},\Cops{4}{\partial^2r^2,3},\Cops{4}{\partial^2rp,3},\Cops{4}{\partial^2\delta^3,2},\Cops{4}{\partial^2r^3,2}\}\nonumber\\
    \Pi\text{-basis:}\hspace{0.3cm}&\{\partial_i\partial_j([\Pi^{[1]}\Pi^{[1]}]_{ij},\Pi_{ij}^{[2]},[\Pi^{[1]}\Pi^{[1]}\Pi^{[1]}]_{ij},[\Pi^{[1]}\Pi^{[1]}]_{ij}\text{Tr}[\Pi^{[1]}],\text{Tr}[\Pi^{[1]}\Pi^{[1]}][\Pi^{[1]}]_{ij},\nonumber\\
    &\text{Tr}[\Pi^{[1]}]\Pi_{ij}^{[2]},[\Pi^{[2]}\Pi^{[1]}]_{ij},[\Pi^{[3]}]_{ij},\text{Tr}[\Pi^{[2]}]\Pi^{[1]}_{ij},\Pi^{[1]}_{ij}(\text{Tr}[\Pi^{[1]}])^3,[(\Pi^{[1]})^4]_{ij},\nonumber\\
    &[(\Pi^{[1]})^3]_{ij}\text{Tr}[\Pi^{[1]}],\text{Tr}[(\Pi^{[1]})^3]\Pi_{ij}^{(1)},[\Pi^{(1)}\Pi^{(1)}]_{ij}(\text{Tr}[\Pi^{[1]}])^2,\nonumber\\
    &[\Pi^{[2]}\Pi^{[1]}\Pi^{[1]}]_{ij},\text{Tr}[(\Pi^{[1]})^2]\Pi_{ij}^{[2]}),\nonumber\\
    &\partial^2(\text{Tr}[\Pi^{[1]}],\text{Tr}[\Pi^{[1]}]^2,\text{Tr}[\Pi^{[1]}\Pi^{[1]}],\text{Tr}[\Pi^{[2]}],\text{Tr}[\Pi^{[1]}]^3,\text{Tr}[\Pi^{[1]}\Pi^{[1]}\Pi^{[1]}],\nonumber\\
    &\text{Tr}[\Pi^{[1]}\Pi^{[1]}]\text{Tr}[\Pi^{[1]}],\text{Tr}[\Pi^{[2]}\Pi^{[1]}],\text{Tr}[\Pi^{[2]}]\text{Tr}[\Pi^{[1]}],\text{Tr}[\Pi^{[3]}],\text{Tr}[\Pi^{[1]}]^4,\text{Tr}[(\Pi^{[1]})^4],\nonumber\\
    &\text{Tr}[(\Pi^{[1]})^3]\text{Tr}[\Pi^{[1]}],\text{Tr}[(\Pi^{[1]})^2]\text{Tr}[\Pi^{[1]}]^2,\text{Tr}[\Pi^{[2]}\Pi^{[2]}])\}\;.
\end{align}
We have checked explicitly that time non-local and $\Pi$ basis as given in Eq.\eqref{eq:scala_op_4} are related by an invertible linear map. The extra time non-local operator appearing Eq.\eqref{eq:scala_op_4} will affect the renormalised velocity divergence ($\theta$) as given in Eq.\eqref{eq:fluid_eq_long}.

\section{Comments on Renormalisation}\label{ap:renorm}
In a generic bias expansion for tracers given in Eq.\eqref{eq:tracer_loc_exp}, we can see that we have product of operators evaluated at the same point in space. This leads to divergences when computing an observable at higher order. To remove these divergences we need to renormalise such operators by adding appropriate counterterms. This needs to be done with all operators to ensure all observables are finite. 

In this section, we briefly discuss how all the selection operators are generated as counterterms by renormalisation of the flux field $\delta_F$ in Lyman-$\alpha$. This is the "top-down" approach to get the same bias expansion as given in Eq.\eqref{eq:bias_exp_selec}, as opposed to the "bottom-up" approach that we have discussed in Sec.\,\ref{ssec:sel_op}. Along the way we will review some of the equation relevant in the discussion of Lyman alpha forest. 

For the case of Lyman alpha forest, we measure flux of photons, $F$ coming towards our line of sight. The fluctuation over the mean value of the flux field, $\Bar{F}$, is defined as,
\begin{align}\label{eq:delF_def}
    \delta_F=\frac{F-\Bar{F}}{\Bar{F}}\;.
\end{align}

The total flux $F$ received depends on the optical depth ($\tau$) of neutral hydrogen along the line of sight of observation in an inverse manner. Optical depth is proportional to the number density of neutral hydrogen. So if the number density is high then there will be more absorption and hence flux will be reduced. The number density of neutral hydrogen is a direct tracer of the underlying dark matter. Being a direct tracer of the underlying dark matter, the optical depth $\delta_\tau$ which is proportional to the number density has the same bias expansion as given in Eq.\eqref{eq:tracer_loc_exp}. Mathematically, the total flux $F$ is related to the optical depth by an exponential map as follows\,\cite{Gunn:1965hd,Croft:1997jf},
\begin{align}\label{eq:exp_map}
    F=e^{-\tau_0(1+\delta_\tau^r)}\;,
\end{align}
where $\tau_0$ is the mean optical depth and $\delta_\tau^r$ is the fluctuation in the optical depth in redshift space. 
Now, using Eq.\eqref{eq:exp_map}, we can write how the flux fluctuation in the Lyman alpha field is related to the fluctuation in the optical depth which is given as\,\cite{Ivanov:2023yla},
\begin{align}\label{eq:delF_expan}
    \delta_F=-\tau_0\delta_\tau^r+\frac{1}{2}\tau_0^2(\delta_\tau^{2,r}-\langle\delta_\tau^{2,r}\rangle)-\frac{1}{6}\tau_0^3\delta_\tau^{3,r}\;.
\end{align}
In order to get an expansion of flux fluctuation field $\delta_F$, we should have the expression for optical depth fluctuation in redshift space.  

So, let's look at the structure of operators that arise when we go to redshift space. Redshift space is related to the real space by a coordinate transformation given as\,\cite{Matsubara:2007wj,Senatore:2014vja},
\begin{align}\label{eq:coord_trans}
    \xvec_r=\xvec+\frac{\vec{v}\cdot\hat{z}}{aH}\hat{z}\;,
\end{align}
where $\xvec_r$ and $\xvec$ are the redshift space and real coordinates respectively and $\hat{z}$ is the unit vector along the line of sight. Under the coordinate change given in Eq.\eqref{eq:coord_trans} and imposing mass conservation, the optical depth fluctuation is expressed in momentum space as,
\begin{align}\label{eq:delta_g_rs}
    \delta_\tau^r(\boldk)=\delta_\tau(\boldk)+\int d^3xe^{-i\boldk\cdot\boldx}\Bigg(\Bigg[\exp{-i\frac{\vec{k}\cdot\hat{z}}{aH}\vec{v}(\xvec)\cdot\hat{z}}-1\Bigg](1+\delta_\tau(\xvec))\Bigg)\;,
\end{align}
where $\delta_\tau(\vec{k})$ has the same bias expansion as given in Eq.\eqref{eq:tracer_loc_exp}. We can expand Eq.\eqref{eq:delta_g_rs} up to desired order in overdensity to get the bias expansion in redshift space. For example up to third order, we have,
\begin{align}\label{eq:rs_expanded}
    \delta_\tau^r(\boldk)&=\delta_\tau(\boldk)-\frac{ik_z}{aH}[v_z]_{\boldk}+\frac{1}{2}\Bigg(\frac{ik_z}{aH}\Bigg)^2[v_z^2]_{\boldk}-\frac{1}{6}\Bigg(\frac{ik_z}{aH}\Bigg)^3[v_z^3]_{\boldk}\nonumber\\
    &-\frac{ik_z}{aH}[v_z\delta_\tau]_{\boldk}+\frac{1}{2}\Bigg(\frac{ik_z}{aH}\Bigg)^2[v_z^2\delta_\tau]_{\boldk}\;,
\end{align}
where $[f]_{\boldk}\equiv \int d^3xe^{-i\boldk\cdot\boldx}f(\xvec)$ and we have used $v_z\equiv \vec{v}.\hat{z}, k_z\equiv \vec{k}.\hat{z} $ for brevity. From Eq.\eqref{eq:rs_expanded}, we can see that a subset of selection operators are generated by going to redshift space.

We can do a time non-local expansion of Eq.\eqref{eq:rs_expanded} to get the set of non-local operators in redshift space. We have checked that up to third order, the non-local operators can be written in terms of the local operators. Hence, time non-locality does not appear in redshift space up to third order.

We can substitute the expression for $\delta_\tau^r$ from Eq.\eqref{eq:rs_expanded} in Eq.\eqref{eq:delF_expan} to get the expression for flux fluctuation in redshift space. The second and third terms in Eq.\eqref{eq:delF_expan} are composite operators. Therefore, it has been pointed out that the second and the third term in Eq.\eqref{eq:delF_expan} need to be renormalised\,\cite{Ivanov:2023yla} appropriately. Upon renormalisation we expect that all operators allowed by the symmetry are generated as counterterms. Therefore,  we get the same bias expansion as in Eq.\eqref{eq:loc_sel_op_3}. To generate all selection operators upto third order, we need third order tree level counterterms. The systematic method of renormalisation for local Eulerian halo biasing in the context of $SO(3)$ invariant operators is developed in\,\cite{Assassi:2014fva}. From that study, it can be observed that to generate all counterterm at third order, we need to evaluate the quantity,
\begin{align}
\langle\delta_{F,\boldsymbol{q}}^{(5)}\delta_{\boldsymbol{q}_{1}}^{(1)}\delta_{\boldsymbol{q}_{2}}^{(1)}\delta_{\boldsymbol{q}_{3}}^{(1)}\rangle\;,
\end{align}
where $\delta_{\boldsymbol{q}_{1}}^{(1)}$ are linear overdensity in dark matter.
Now, $\delta_F$ has the expansion given in Eq.\eqref{eq:delF_expan} which contains $\delta_\tau^r$. The kernel for $\delta_F$ upto fifth order in momentum space is too big to deal with. Furthermore, identifying individual selection operators is impractical, therefore we leave that exercise for future and assume that all selection operators are generated when $\delta_F$ is renormalised to all loops.

\bibliographystyle{JHEP.bst}
\bibliography{references} 

\end{document}